%% file: paper_clean.tex
\documentclass{aa}
\usepackage{textcomp}
\usepackage{graphicx} %
\usepackage{amsmath} %
\usepackage{amssymb} %
\usepackage{bm} %
\usepackage{upgreek} %
\usepackage{IEEEtrantools} %
\usepackage{multirow}
\usepackage[colorlinks=true,allcolors=blue,urlcolor=blue]{hyperref}
\usepackage{txfonts}
\usepackage[normalem]{ulem} 

\newcommand{\sect}[1]{\text{Sect.~\ref{#1}}}
\newcommand{\fig}[1]{\text{Fig.~\ref{#1}}}
\newcommand{\tab}[1]{\text{Table~\ref{#1}}}

\newcommand{\multitd}{\texttt{Multi3D}}
\newcommand{\balder}{\texttt{Balder}}

\newcommand{\marcs}{\texttt{MARCS}}
\newcommand{\stagger}{\texttt{Stagger}}
\newcommand{\atmo}{\texttt{ATMO}}

\newcommand{\lgeps}[1]{A\mathrm{\left(#1\right)}}
\newcommand{\eV}{\mathrm{eV}}
\newcommand{\dex}{\mathrm{dex}}
\newcommand{\nm}{\mathrm{nm}}

\newcommand{\kelvin}{\mathrm{K}}
\newcommand{\kms}{\mathrm{km\,s^{-1}}}
\newcommand{\teff}{T_{\mathrm{eff}}}
\newcommand{\lgg}{\log{g}}
\newcommand{\feh}{\ensuremath{\mathrm{[Fe/H]}}}
\newcommand{\vmic}{\xi}
\newcommand{\xfe}[1]{\ensuremath{\mathrm{[#1/Fe]}}}
\newcommand{\xh}[1]{\ensuremath{\mathrm{[#1/H]}}}
\newcommand{\lggf}{\log g f}

\begin{document} 

\title{A distinct halo population revealed from 3D non-LTE magnesium abundances}
\author{T. Matsuno\inst{\ref{ari},\ref{groningen}}
\and
A.~M.~Amarsi\inst{\ref{uu1}}
\and
M.~Carlos~\inst{\ref{uu1}}
\and
P.~E.~Nissen\inst{\ref{au1}}}
\institute{
\label{ari}Astronomisches Rechen-Institut, Zentrum f\"ur Astronomie der Universit\"at Heidelberg, M\"onchhofstra{\ss}e 12-14, 69120 Heidelberg, Germany \email{matsuno@uni-heideberg.de}
\and
\label{groningen}Kapteyn Astronomical Institute, University of Groningen, Landleven 12, 9747 AD Groningen, The Netherlands
\and
\label{uu1}Theoretical Astrophysics, 
Department of Physics and Astronomy,
Uppsala University, Box 516, SE-751 20 Uppsala, Sweden \email{anish.amarsi@physics.uu.se}
\and
\label{au1}Department of Physics and Astronomy,
Aarhus University, Ny Munkegade 120, DK-8000 Aarhus C,
Denmark}

\abstract{
Magnesium is one of the important elements in stellar physics as an electron donor and in Galactic Archaeology as a discriminator of different stellar populations. 
However, previous studies of \ion{Mg}{I} and \ion{Mg}{II} lines in metal-poor benchmark stars 
have flagged problems with magnesium abundances inferred from one-dimensional (1D), hydrostatic models of stellar atmospheres, both with or without the local thermodynamic equilibrium (LTE) approximation.
We here present 3D non-LTE calculations for magnesium in FG-type dwarfs, and provide corrections for 1D LTE abundances. 
The 3D non-LTE corrections reduce the ionisation imbalances
in the benchmark metal-poor stars HD84937 and HD140283
from $-0.16\,\dex$ and $-0.27\,\dex$ in 1D LTE, to just
$-0.02\,\dex$ and $-0.09\,\dex$ respectively.
We then applied our abundance corrections to 1D LTE literature results for stars in the thin disc, thick disc, $\alpha$-rich halo, and $\alpha$-poor halo.
We find that the 3D non-LTE results show a richer substructure in $\xfe{Mg}-\xh{Fe}$ in the $\alpha$-poor halo, revealing two subpopulations at the metal-rich end.
These two subpopulations are also separated in kinematics, supporting the astrophysical origin of the separation.
While the more magnesium-poor subpopulation is likely to be debris from a massive accreted galaxy, Gaia-Enceladus, the other subpopulation may be related to a previous identified group of stars, called Eos.
The presence of additional separation in $\xfe{Mg}$ suggests that previous Mg abundance measurements may have been limited in the precision by the 1D and LTE approximations, highlighting the importance of 3D non-LTE modelling.
}


\date{Received / Accepted}
\maketitle

\section{Introduction}
\label{introduction}

Magnesium is a highly important element in both stellar and galactic
astrophysics. Owing to its relatively high cosmic abundance
($\lgeps{Mg}=7.55$ in the
Sun; \citealt{2021A&A...653A.141A}) and low ionisation potential
($E_{\text{ion.}}= 7.65\,\eV$), magnesium is the main electron donor in the
photosphere of the Sun and similar stars.\footnote{$\lgeps{Mg}=\log_{10}\frac{N_{\mathrm{Mg}}}{N_{\mathrm{H}}}+12$.}  Magnesium abundances measured in
stars have in recent years served as a key tracer of Galactic chemical
evolution \citep{1998A&A...338..161F,2019ApJ...874..102W},
benefiting from data for hundreds of thousands of stars coming from 
the APOGEE \citep{2022ApJS..259...35A}
and GALAH \citep{2021MNRAS.506..150B} surveys.
In particular, \citet{2010A&A...511L..10N}
demonstrated that the thick disc and inner halo of the Milky Way separates
into two components with high and low magnesium abundances.
In light of results from Gaia DR2
\citep{2018MNRAS.478..611B,2018Natur.563...85H},
the latter (low) component is interpreted as an accreted population of stars,
illustrating how magnesium abudances
may serve as a diagnostic for Galactic Archaeology.
Now that multiple kinematic substructures are known to differ in magnesium abundance \citep{2022A&A...661A.103M,2022A&A...665A..46M,2023MNRAS.520.5671H}, one of the next questions is to how many components we can clearly separate halo stars just using chemical abundance.
Can the 1D LTE approximations widely used be a bottleneck in such approaches?

Accurate magnesium abundance determinations are therefore of great interest to
the stellar and Galactic astrophysics communities.
One source of systematic error that is frequently discussed in the
literature arise from assumptions made about the stellar atmosphere ---
in particular, that they are one dimensional (1D) and hydrostatic,
and that the medium satisfies local thermodynamic equilibrium (LTE).
Several studies based on 1D models have already illustrated potentially large
departures from LTE in \ion{Mg}{I} lines, 
primarily driven by overionisation of the minority neutral
species \citep[e.g.][]{2015A&A...579A..53O,
2018ApJ...866..153A,2022A&A...665A..33L}, increasing inferred
magnesium abundances relative to 1D LTE.
In their 1D non-LTE studies,
\citet{2018ApJ...866..153A} and \citet{2022A&A...665A..33L}
draw attention to significant ionisation imbalances present
in the benchmark metal-poor F-dwarf HD84937 
and G-subgiant HD140283,
where magnesium abundances inferred from
the neutral species are typically around $0.2\,\dex$ lower than those
inferred from the singly-ionised species.
They suggest that 3D non-LTE effects
could help solve this problem.
To date, 3D non-LTE effects have only been quantified for the Sun
by \citet{2021A&A...653A.141A}, where the corrections indeed go
in the right direction,
increasing magnesium abundances inferred from the neutral species
relative to 1D non-LTE; and for two theoretical
models by \citet{2017ApJ...847...15B}, where the corrections are strongly
dependent on the adopted microturbulence (a fudge parameter introduced in 1D
models to account for 3D effects).

We here present the results of
3D non-LTE calculations for a subset of 
$40$ 3D radiative-hydrodynamic simulations in the \stagger{}-grid
\citep{2013A&A...557A..26M} covering
FG-type dwarfs (\sect{method}).
To illustrate their possible 
impact and validate the data, we inspect the ionisation balance in 
HD84937 and HD140283 (\sect{benchmark}).
We then reanalyse the \ion{Mg}{I} $571.1\,\nm$ line in a sample of disc and halo stars,
in particular the $\alpha$-poor and $\alpha$-rich halo populations
identified in \citet{2010A&A...511L..10N}
where they help to reveal substructures in abundance space, that may
be related to separate acccretion events in the Milky Way's history
(\sect{halo}).
We make the corrections available to the community
in electronic format (\sect{conclusion}).

\section{Method}
\label{method}

\input{linelist.tex}

\input{abundances.tex}

\subsection{3D non-LTE calculations}
\label{method-calc}

The post-processing calculations were carried out with \balder{}
\citep{2018A&A...615A.139A}, a 3D non-LTE code with roots in \multitd{}
\citep{2009ASPC..415...87L_short} but with updates in particular
to the equation of state and opacity package
\citep{2023A&A...677A..98Z}.
The method of calculation follows that described in 
\citet{2022A&A...668A..68A}: in particular,
Rayleigh scattering from hydrogen was included, and other background
species were treated in pure absorption.
The model atom for neutral and singly-ionised magnesium 
was described in \citet{2021A&A...653A.141A}.

Calculations were performed for a suite of 3D and 1D model atmospheres.
The 3D model atmospheres come from the \stagger{}-grid
\citep{2013A&A...557A..26M}. The $40$ models span
$5000\,\kelvin\lesssim\teff\lesssim6500\,\kelvin$ in steps of approximately
$500\,\kelvin$; $\lgg$ of $4.0\,\dex$ and $4.5\,\dex$ for all models, and also
$5.0\,\dex$ for models with $\teff$ close to $5000\,\kelvin$ and
$5500\,\kelvin$; and $-3.0\leq\feh\leq0.0$ in steps of $1.0\,\dex$.
The model atmospheres adopt solar abundances
from \citet{2009ARA&A..47..481A} scaled by $\feh$, 
with an enhancement to $\alpha$-elements
$\xfe{\alpha}=+0.4$ for $\feh\leq-1.0$. 
Calculations were also performed on the 1D equivalent of the \stagger{} models
(\atmo{}; see the Appendix of \citealt{2013A&A...557A..26M}), having the
same chemical composition, same $\teff$, and adopting the same opacity binning scheme.
The calculations on 1D \atmo{} models were performed for three different values
of microturbulence ($0$, $1$, and $2\,\kms$), amounting
to $40\times3=120$ calculations in total.
For the spectrum synthesis, the magnesium abundance was kept strictly equal
to that for which the model atmosphere was calculated.

To aid the analysis of HD84937 and HD140283 (\sect{benchmark}), 
1D LTE equivalent widths were calculated
across a more extended and finer grid 
of standard \marcs{} model atmospheres \citep{2008A&A...486..951G}.
The models used here are a
subset of the $3756$ models used in \citet{2020A&A...642A..62A},
spanning $5000\,\kelvin\lesssim\teff\lesssim7000\,\kelvin$ in steps of approximately
$250\,\kelvin$; $3.0\leq\lgg\leq5.0$ in steps of $0.5\,\dex$, and a wide range
of $\feh$. 
They use solar abundances from \citet{2007coma.book..105G} scaled by $\feh$, 
with an enhancement to $\alpha$-elements
$\xfe{\alpha}=+0.4$ for $\feh\leq-1.0$.
Although the $\alpha$-enhancement is fixed for the model atmospheres,
the magnesium abundances were allowed to vary in the 
post-processing line synthesis:
$\xfe{Mg}=-1.0$ to $+1.0$ in steps of $0.2\,\dex$.

Theoretical equivalent widths were derived by direct integration across
the normalised line profiles.  
Abundance corrections relative to 1D LTE (``1N-1L'' for 1D non-LTE,
and ``3N-1L'' for 3D non-LTE) were calculated based on these equivalent widths,
as a function of $\teff$, $\lgg$, $\xh{Mg}$, and the 1D LTE microturbulence $\vmic$.
The abundance corrections were interpolated onto the
parameters of HD84937 and HD140283 in \sect{benchmark}.
For the reanalysis of literature data in
\sect{halo}, differential abundance corrections were applied, by also calculating the abundance corrections for the Sun (see e.g.~\citet{2019A&A...630A.104A}).
In all cases, extrapolation was not permitted; 
the edge values were adopted for the few stars with surface gravities 
lying slightly outside of the \stagger{} and \atmo{} grids, and for 
the disc stars with $\feh>0$.

\subsection{Line parameters and equivalent widths}
\label{method-lines}

In \tab{tab:linelist} we show the lines
that were considered in this study.
The lines are those that were measured by 
\citet{2018ApJ...866..153A} in HD84937 and HD140283.
We also included the \ion{Mg}{I} $571.1\,\nm$ line: the reanalysis of disc
and halo stars in \sect{halo} is based on this line alone.

The table indicates the line parameters adopted in the abundance analysis of HD84937
and HD140283.  In the radiative transfer calculations themselves, the oscillator strengths for \ion{Mg}{I} lines all come from \citet{2017A&A...598A.102P}, with preference given to their experimental values where available. However, for two lines, our abundance results were adjusted to incorporate different data, via $\Delta \lgeps{Mg}=-\Delta\lggf$ which is valid in the weak part of the curve of growth.

First, for the \ion{Mg}{I} $457.1\,\nm$, we speculate that the value of $-5.397$ for the $\lggf$ adopted in our 3D non-LTE calculations is probably systematically too high.  We instead adopted the result from the multiconfiguration Dirac-Fock calculations of \citet{1997APS..APR.J1537J}, $-5.732$, which is in almost perfect agreement with the experimental value of \citet{1992PhRvA..45.1717G} based on laser spectroscopy.\footnote{Their Table 6 result, labelled $\Delta\mathrm{fgh}$.} Our abundance results 
in \sect{benchmark} have thus been increased by $0.33\,\dex$ due to this offset.

Secondly, for the \ion{Mg}{I} $571.1\,\nm$ line, the abundance analysis is based on the theoretical value of \citet{2017A&A...598A.102P}, namely 
$-1.742$.  This is $0.10\,\dex$ larger than their experimental value of $-1.84\pm0.05$, which is what was adopted in our model atom and radiative transfer calculations.  The value on NIST, also theoretical, is $-1.72$ via \citet{1990JQSRT..43..207C}, with an accuracy grade of ``B'' (around $0.04\,\dex$ uncertainty).  
In the analysis of HD84937 and HD140283 (\sect{benchmark}) we have reduced the $\lgeps{Mg}$ inferred from the \ion{Mg}{I} $571.1\,\nm$ by $0.10\,\dex$. 
In the literature reanalysis (\sect{halo}), the choice of $\lggf$ does not directly impact the results, because the shift in $\lggf$ is cancelled out in the solar-differential abundance corrections.

\section{3D non-LTE effects and ionisation balance in HD84937 and HD140283}
\label{benchmark}

\begin{figure*}
    \begin{center}
        \includegraphics[scale=0.325]{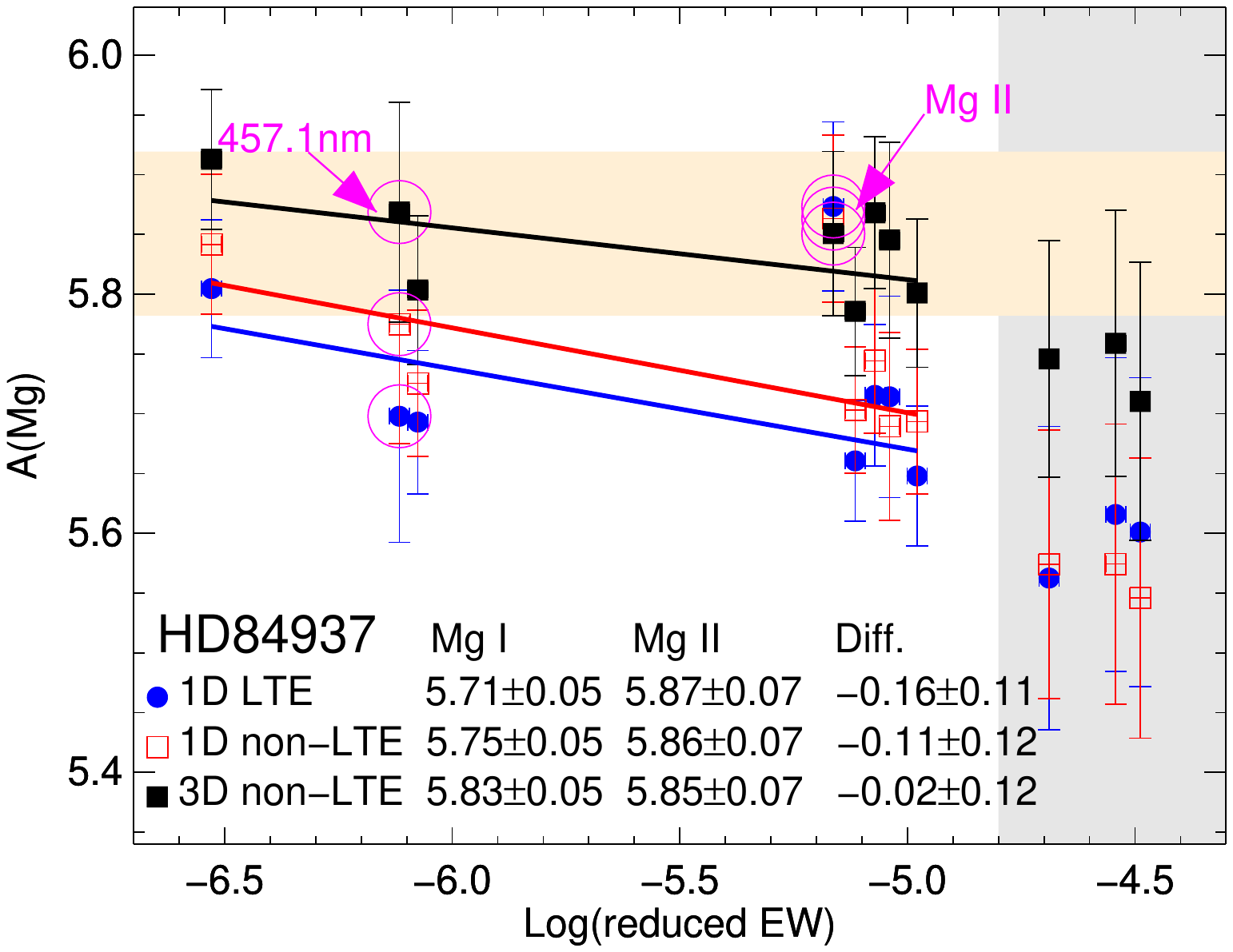}\includegraphics[scale=0.325]{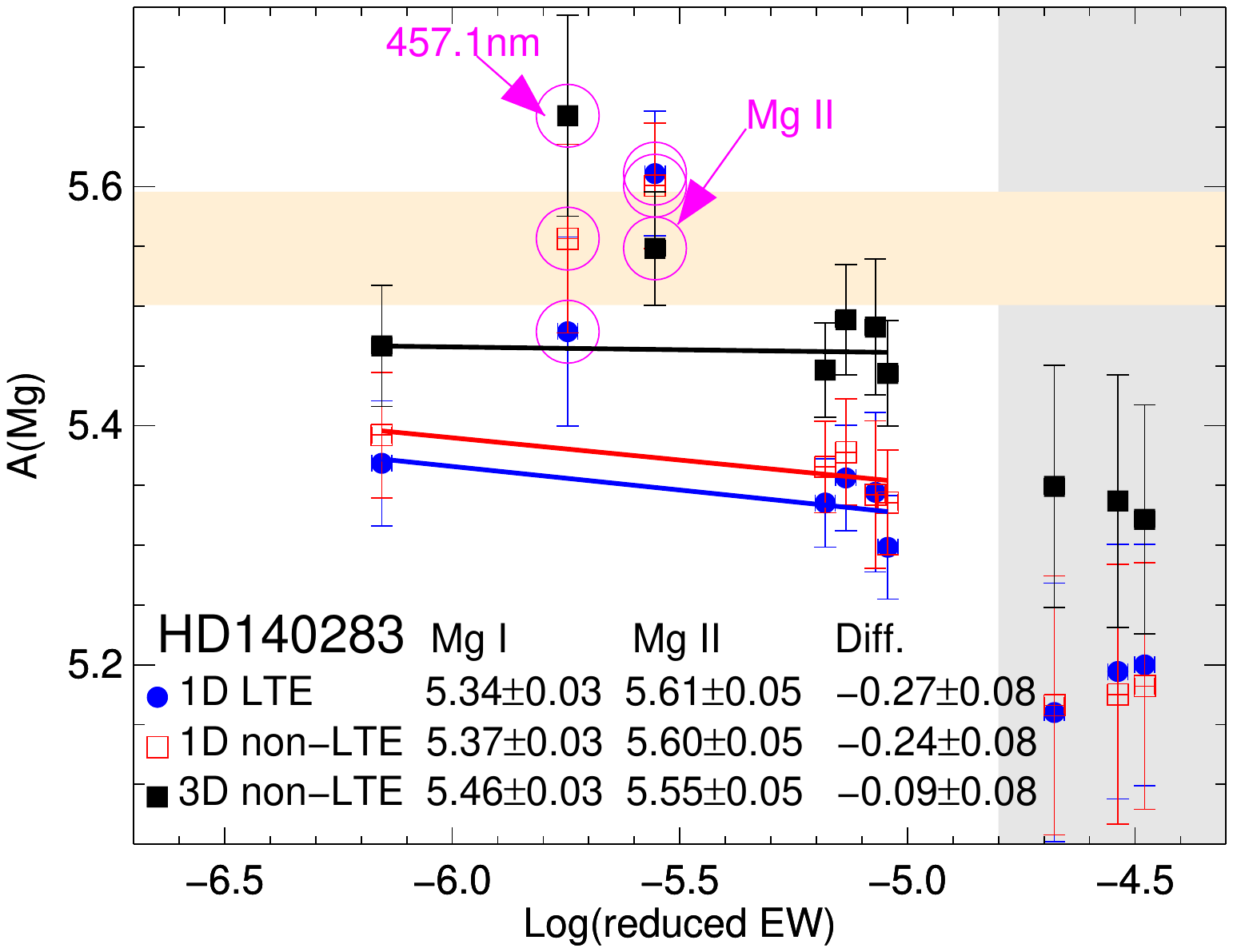}
        \caption{Magnesium abundances
        for HD84937 (left) and HD140283 (right), based on line-by-line
        equivalent widths.
        Error bars reflect uncertainties in $\teff$, $\lgg$, $\feh$,
        and the $\vmic$ adopted in the 1D analysis, as well
        as a $5\%$ uncertainty in the equivalent width
        ($10\%$ uncertainty adopted 
        for the weak \ion{Mg}{I} $473.0\,\nm$ line in HD84937
        and for the blended \ion{Mg}{II} $448.1\,\nm$ line).
        Weighted lines of best fit to \ion{Mg}{I} lies
        overdrawn, excluding the saturated $517\,\nm$ triplet
        and the $457.1\,\nm$ line that forms in the upper layers of the photosphere.
        Shaded horizontal area gives the 3D non-LTE result and uncertainty
        from the \ion{Mg}{II} $448.1\,\nm$ line. 
        The weighted mean abundances and the abundance difference, and their uncertainties indicated in the legend are computed using the framework of \citet{2020AJ....160..181J}.}
        \label{fig:benchmark}
    \end{center}
\end{figure*}

\tab{tab:abundances} shows the adopted equivalent widths and the results of the 
abundance analysis of HD84937 and HD140283.
For most lines the measured equivalent widths are from \citet{2022A&A...665A..33L}.
The only exceptions are the \ion{Mg}{I} $416.7\,\nm$ and the \ion{Mg}{II} $448.1\,\nm$.
For the \ion{Mg}{I} $416.7\,\nm$ in HD84937, a value of $3.20\,\mathrm{pm}$ was adopted, instead of $3.65\,\mathrm{pm}$, 
after accounting for weak blends in the wings of the line.
For the \ion{Mg}{II} $448.1\,\nm$ the impact of the \ion{Ti}{I} blend
\citep[e.g.][]{2018ApJ...866..153A} was estimated in the following way. We measured $\lgeps{Ti}=3.21$ and $2.71$ for the two stars respectively based on a 1D LTE analysis of \ion{Ti}{I} and \ion{Ti}{II} lines; these values agree well with the 1D LTE results of \citet{2022A&A...668A.103M}.
Although there are 1D non-LTE effects on \ion{Ti}{I} lines corresponding to positive abundance corrections of the order $0.2\,\dex$,
Mallinson et al. (in press) suggest that 3D effects go in the opposite direction in metal-poor dwarfs: as such, in the absence of full 3D non-LTE calculations for \ion{Ti}{I}, we adopt 1D LTE results.  Taking $\lggf=0.17$ for the blend \citep{2013ApJS..205...11L}, the contribution to the equivalent widths are 
$0.15\,\mathrm{pm}$ and $0.12\,\mathrm{pm}$ in HD84937 and HD140283, respectively.
These values were subtracted from the equivalent widths reported in \citet{2022A&A...665A..33L}.
Using synthetic spectra, we confirmed that this approach does not introduce any significant error. 
The total equivalent widths of synthetic spectra with both \ion{Mg}{II} and \ion{Ti}{I} lines included are reproduced within $0.03\mathrm{pm}$ by adding equivalent widths of the two lines.
The $0.03\mathrm{pm}$ affects Mg abundance by less than 0.01 dex, which is much smaller than the other sources of uncertainties. 
Uncertainties on the equivalent widths of $5\%$ were adopted for most lines:
for the very weak lines (\ion{Mg}{I} $457.1\,\nm$, $473.0\,\nm$, and $571.1\,\nm$)
$10\%$ uncertainties were adopted owing to the uncertainty of placing the continuum.

The stellar parameters and $1\,\sigma$ uncertainties were adopted from the literature.  For HD84937,
$\teff=6356\pm97\,\kelvin$ \citep{2015A&A...582A..49H},
$\lgg=4.13\pm0.03$ \citep{2021A&A...650A.194G}, and $\feh=-1.96\pm0.03$ 
(3D non-LTE value from \citealt{2022A&A...668A..68A}).
For HD140283, 
$\teff=5792\pm55\,\kelvin$ and 
$\lgg=3.65\pm0.02$ \citep{2020A&A...640A..25K},
and $\feh=-2.28\pm0.02$ 
(3D non-LTE value from \citealt{2022A&A...668A..68A}).
For the 1D analyses, $\vmic=1.39\pm0.24\,\kms$ and 
$1.56\pm0.20\,\kms$ were assumed \citep{2015A&A...582A..81J}.

The uncertainties on the line-by-line 
abundances reported in \tab{tab:abundances}
were estimated by sampling from
Gaussian distributions in $\teff$, $\lgg$, $\feh$,
and $\vmic$, with standard deviations from
the $1\,\sigma$ uncertainties given above.
The uncertainties in the equivalent widths were taken into
account in the same way.
Using the framework of \citet{2020AJ....160..181J}, weights are then given to the lines from sensitivities of the derived abundances to the stellar parameters and the equivalent widths and used to compute weighted mean abundances and their uncertainties.
In this approach, the uncertainties are estimated properly; e.g., correlated uncertainties among lines due to their similar sensitivities to stellar parameters are considered.

The columns 3N-1N and 1N-1L give the 3D non-LTE versus 1D non-LTE
and 1D non-LTE versus 1D LTE abundance corrections.
The \ion{Mg}{II} $448.12\,\nm$ line suffers mild 
negative 1N-1L corrections, but the neutral minority species is susceptible to overionisation \citep[e.g.][]{2018ApJ...866..153A}. Thus the weak \ion{Mg}{I} lines feel significant positive 1N-1L corrections.
As the \ion{Mg}{I} lines begin to saturate they become more susceptible to competing photon losses, and in this model there are slightly negative 1N-1L corrections for the \ion{Mg}{I} $880.67\,\nm$ line as well as for the \ion{Mg}{I}  $517\,\nm$ triplet.  Similar to \ion{Fe}{I} \citep{2016MNRAS.463.1518A}, the 3N-1N corrections tend to go in the same direction as the 1N-1L corrections for the \ion{Mg}{I} lines, in part due to the steeper temperature gradients coupling with and enhancing the dominating overionisation effect  \citep[e.g][]{2023A&A...672A..90L}.  This means that in metal-poor stars it is important to model \ion{Mg}{I} lines in full 3D non-LTE.

This point is demonstrated by considering the ionisation balance of HD84937 and HD140283.
To that end,
\fig{fig:benchmark} shows the line-by-line magnesium abundances for the two stars
in 1D LTE, 1D non-LTE, and 3D non-LTE.
We find that ionisation balance is improved in 3D non-LTE, 
compared to 1D LTE and 1D non-LTE, as predicted by
\citet{2018ApJ...866..153A} and \citet{2022A&A...665A..33L}.
To quantify this, we neglect 
the \ion{Mg}{I} $517\,\nm$ triplet which is saturated and thus not a good abundance diagnostic.  We also neglect the
\ion{Mg}{I} $457.1\,\nm$ line, which gives an anomalous result in HD140283;
in the Sun this line forms in the lower chromosphere
\citep[e.g.][]{1975SoPh...42..289A,2009ApJ...696.1892L}.
After neglecting these two lines, the ionisation imbalance is
slightly reduced by $0.03--0.05\,\dex$ when 
going from 1D LTE to 1D non-LTE;
i.e.,~1D LTE gives $-0.16\,\dex$ and $-0.27\,\dex$ for HD84937 and HD140283 respectively (the negative sign indicating that
\ion{Mg}{I} lines give abundances that are too low),
while 1D non-LTE gives $-0.11\,\dex$ and $-0.24\,\dex$.
The improvement when going from 1D LTE or 1D non-LTE to 3D non-LTE is much more significant; the ionisation imbalance is reduced to just $-0.02\,\dex$ and $-0.09\,\dex$ in the two stars respectively.
This suggests that 3D non-LTE effects should be taken into account for the accurate characterisation of warm metal-poor dwarfs and subgiants.

The residual imbalance in HD140283 of $-0.09\,\dex$ is puzzling.  Considering the (correlated) uncertainties on the abundances from \ion{Mg}{I} and \ion{Mg}{II} lines,
the imbalance has an uncertainty of 0.08 dex and a significance of $1\,\sigma$. 
The uncertainty is dominated by that in the effective temperature.  The interferometric value of \citet{2020A&A...640A..25K} was adopted here ($\teff=5792\pm55\,\kelvin$), which is consistent with that inferred from the Infrared Flux Method 
\citep{2010A&A...512A..54C,2018MNRAS.475L..81K} and from the wings of hydrogen lines 
\citep{2018A&A...615A.139A,2021A&A...650A.194G}. Increasing $\teff$ by its $1\,\sigma$ uncertainty would significantly improve the ionisation balance. A second possibility is that the impact of the titanium blend on the \ion{Mg}{II} line is underestimated, noting that the blend affects HD140283 much more strongly that HD84937.  Taking a higher titanium abundance $\lgeps{Ti}=3.09$ (via $\xfe{Ti}\approx0.4$), the ionisation imbalance is reduced to just $-0.02\,\dex$.  A detailed 3D non-LTE abundance analysis of titanium in HD140283 combined with a 3D non-LTE synthesis of the blend could therefore help shed light on this problem. Finally, uncertainties in the non-LTE modelling of \ion{Mg}{II} lines cannot be ruled out. In the solar atmosphere the ionised species was found to be highly sensitive to the hydrogen collisions \citep{2021A&A...653A.141A}.  The 1D non-LTE corrections found here are less severe than those reported by \citet{2018ApJ...866..153A} and \citet{2022A&A...665A..33L} for this star.  Slightly larger departures from 1D LTE would help resolve the discrepancy.  The current model adopts the simplistic Drawin recipe \citep[e.g.][]{1993PhST...47..186L}; improved data based on asymptotic models \citep[e.g.][]{2016PhRvA..93d2705B,2017A&A...606A.147B} would be welcome.

\section{Magnesium abundances of stellar populations}
\label{halo}

\begin{figure*}
    \begin{center}
    \includegraphics[width=\textwidth]{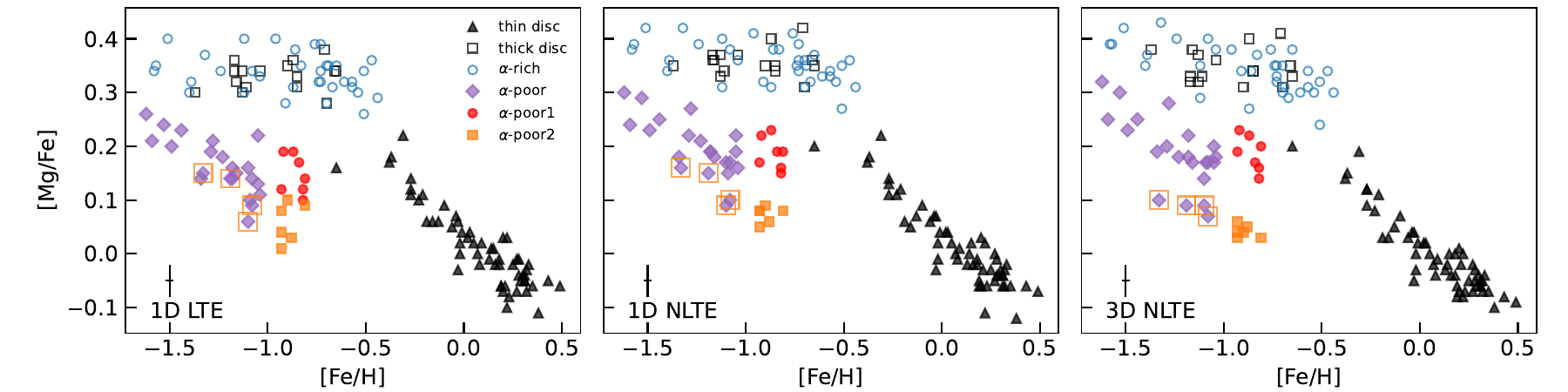}
    \caption{Magnesium abundances for the samples of \citet{2010A&A...511L..10N} and Carlos et al. (in prep.) in 1D LTE, 1D non-LTE, and in 3D non-LTE. In all panels the Fe abundance is based on a 3D LTE analysis of \ion{Fe}{II} lines \citep{2019A&A...630A.104A}. The metal-rich end ($-1<\feh$) of the $\alpha$-poor population is further divided into two, $\alpha$-poor 1 \& 2, based on 3D non-LTE $\xfe{Mg}$ ratios. We marked four metal-poor $\alpha$-poor stars with open squares as they have slightly lower \xfe{Mg} in 3D non-LTE than the other $\alpha$-poor stars at the same metallicity.}
    \label{fig:MgFe}
    \end{center}
\end{figure*}

We here investigate the possible impact of 3D non-LTE magnesium abundance in studying stellar populations in the Milky Way, using the sample of halo stars in \citet{2010A&A...511L..10N} and the sample of thin disc stars analysed in Carlos et al. (in prep.).\footnote{We note that an earlier version of our 3D non-LTE corrections was applied to the sample of \citet{2010A&A...511L..10N} in \citet{2024A&A...682A.116N}.} 
All the stars studied in the current  work presents $\teff$ and $\vmic$ from \citet{2014A&A...568A..25N}, while $\feh$ values  were determined by \citet{2019A&A...630A.104A}, $\lgg$ for the halo stars are also from \citet{2014A&A...568A..25N} and new $\lgg$ were calculated adopting GAIA DR3 data \citep{2023A&A...674A..32B,2023A&A...674A...1G} for the thin disc sample.
The present analysis is based on 1D non-LTE and 3D non-LTE solar-differential corrections to the 
1D LTE $\xh{Mg}$ from the above papers, considering the
\ion{Mg}{I} $571.1\,\nm$ line alone.

\fig{fig:MgFe} shows the results of our reanalysis.  The differential corrections to the solar-normalised $\xfe{Mg}$ are more severe at lower metallicities, as expected because these stars are all dwarfs with effective temperatures within several hundred kelvin from the Sun. We therefore subsequently focus on the halo stars.  The original analysis of \citet{2010A&A...511L..10N} revealed the existence of two distinct stellar populations among halo stars based on precise magnesium abundance with a typical uncertainty of 0.03-0.04 dex measured from careful analysis of high-resolution, high-$S/N$ spectra. 
While their analysis was in 1D LTE, our 3D non-LTE corrections now allow us to explore stellar populations in this sample with accurate and even more precise magnesium abundances.
This will enable us to investigate if there are further subpopulations, especially among the $\alpha$-poor, accreted population, since we know more than one galaxy has been accreted to the Milky Way.  
The population indeed shows a larger scatter in elemental abundance ratios \citep{2011A&A...530A..15N}, and kinematically selected subsets seem to have slightly different abundance ratios compared to the rest of the $\alpha$-poor population (see Figures~5-9 of 
\citealt{2022A&A...661A.103M}).

\begin{figure}
    \begin{center}
    \includegraphics[width=0.5\textwidth]{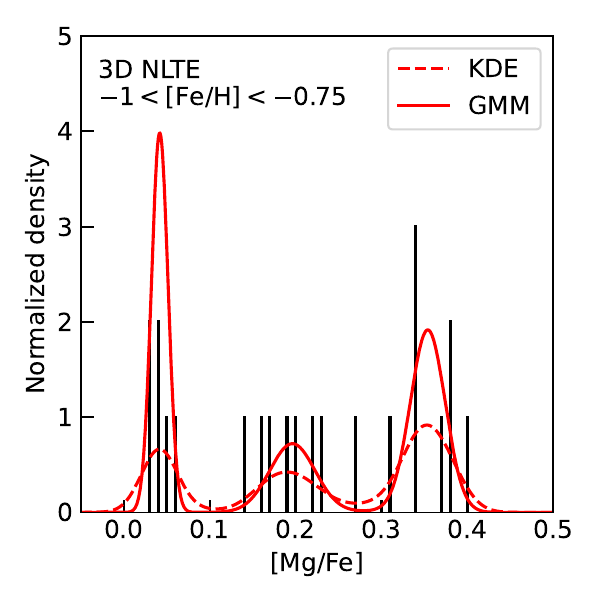}
    \caption{The \xfe{Mg} distribution of metal-rich halo stars with $-1<\feh$. The red solid line shows the estimated density based on a Gaussian Mixture model (GMM) with the number of components of 3, and the dashed line shows that based on a Gaussian Kernel Density Estimation (KDE) with the bandwidths of 0.03 dex. Three peaks, corresponding to the $\alpha$-rich population and the two $\alpha$-poor subpopulations, are clearly visible in both density estimations.}
    \label{fig:MgFe1D}
    \end{center}
\end{figure}

As we move from 1D LTE to 1D non-LTE and 3D non-LTE analysis results, while the overall separation between $\alpha$-rich and poor populations remains present, there emerges another separation at the metal-rich end ($-1<\feh$) of the $\alpha$-poor population. 
\fig{fig:MgFe1D} demonstrates the clear separations among the $\alpha$-rich and the two $\alpha$-poor populations at $-1<\feh$. 
We also provide further evidence for the existence of the three distinct populations in Appendix \ref{appendix:A}.
The clearer separation between the two $\alpha$-poor subpopulations is mainly due to the reduced $\xfe{Mg}$ dispersion among each subpopulation; while the differences in average $\xfe{Mg}$ ratios are 0.10, 0.11, and 0.14 dex in 1D LTE, 1D non-LTE, and 3D non-LTE, the scatters are reduced from 0.034 dex to 0.028 and 0.030 dex for the higher $\xfe{Mg}$ $\alpha$-poor population ($\alpha$-poor 1) and 0.033 dex to 0.014 and 0.011 dex for the lower $\xfe{Mg}$ $\alpha$-poor population ($\alpha$-poor 2).
This suggests that the precision was limited by the approximations of 1D and LTE before, and the 3D non-LTE analysis can help to fully utilise the high signal-to-noise ratio of the spectra.
Even for the small ranges of stellar parameters of the sample of \citet{2010A&A...511L..10N}, namely, ($5297\,\mathrm{K}<T_{\rm eff}<6445\,\mathrm{K}$ and $3.77<\log g<4.63$), 1D non-LTE -- 1D LTE correction varies from $-0.01$ to $0.06$ with a dispersion of $0.02$ dex, and 3D non-LTE -- 1D non-LTE varies from $-0.06$ to $0.03$ with a similar dispersion. In total, 3D non-LTE -- 1D LTE correction varies from $-0.06$ to $0.08$ with a dispersion of $0.04$ dex, which is comparable to or slightly larger than the reported uncertainty on [{Mg}/{Fe}] \citet{2010A&A...511L..10N}.

\begin{figure*}
    \begin{center}
    \includegraphics[width=\textwidth]{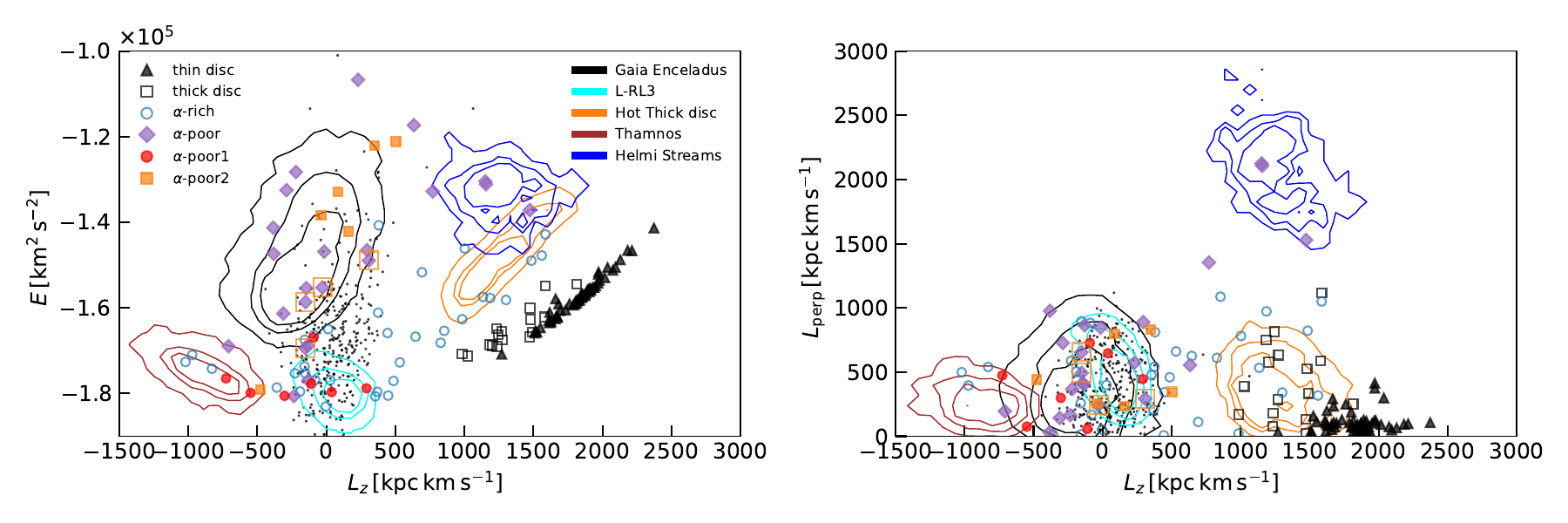}
    \caption{Kinematics of subpopulations. Angular momentum and orbital energy were computed in the same manner as \citet{2022A&A...661A.103M}, and $L_{\rm perp}$ is defined as $L_{\rm perp}=\sqrt{L_x^2 + L_y^2}$. Symbols follow that of \fig{fig:MgFe}. We also show distributions of member stars of major substructures in  \citet{2023A&A...670L...2D} with contours and Eos members from \citet{2022ApJ...938...21M} as black dots. For each substructure, contours encompass 50, 80 and 98\% of its members. Note that the distribution of Eos members is biased since \citet{2022ApJ...938...21M} only used stars with high eccentricity. }
    \label{fig:kinematics}
    \end{center}
\end{figure*}
Interestingly, the $\alpha$-poor 1 \& 2 populations also differ in kinematics (\fig{fig:kinematics}).
Whereas the $\alpha$-poor 1 population stars are preferentially more tightly bound to the Milky Way with low orbital energy with the average (scatter) $\langle E \rangle =(-1.77\pm 0.04)\times 10^5 \,\mathrm{km^2\,s^{-2}}$, the $\alpha$-poor 2 population stars tend to have high orbital energy with $\langle E \rangle = (-1.39\pm 0.19)\times 10^5 \,\mathrm{km^2\,s^{-2}}$.
The kinematics of the $\alpha$-poor 2 population are consistent with being part of Gaia-Enceladus, suggesting that they trace the chemical evolution of the progenitor galaxy of Gaia-Enceladus.
The $\alpha$-poor 1 population partially overlaps in kinematics with L-RL3. 
However, \citet{2022A&A...665A..58R} and \citet{2023A&A...670L...2D} found its chemistry to be a mixture of hot thick-disc and Gaia-Enceladus, which is not consistent with the chemical property of our $\alpha$-poor 1 population.
A similar separation was already noted in \citet{2024A&A...682A.116N}, who also showed that the two separations are different in \xfe{Na} (see also Appendix~\ref{appendix:A}).  

There are outlying populations that deviate from the above overall interpretations and additional stars at lower metallicity with $\feh<-1$ that need discussion.
While G56-30 is chemically classified to $\alpha$-poor 2 population with $\feh=-0.88$ and $\xfe{Mg}_{\rm 3D NLTE}=0.05$, it has a low orbital energy ($E=-1.79\times 10^5\,\mathrm{km^2\,s^{-2}}$, comparable to $\alpha$-poor 1 population and the \xfe{Na} ratio in-between the two populations.
It needs to be seen with a larger sample with precise and accurate chemical abundance if there are stars similar to G56-30 and if they can be considered to be part of Gaia-Enceladus. 
Another group of stars comprises of HD~163810, G176--53, HD~193901, and G21--22, which have $\feh<-1$ and seem to have lower \xfe{Mg} than the other stars at the same metallicity. 
These stars are marked with open orange squares in Figures~\ref{fig:MgFe} and \ref{fig:kinematics}. 
They stand out much less in other elemental abundance ratios, including \xfe{Na}, and their kinematics are similar to other $\alpha$-poor stars, following the distribution of Gaia-Enceladus.
Hence, we consider these four stars part of Gaia-Enceladus, although future studies with a larger sample will be welcomed.

We here further discuss connections of the $\alpha$-poor 1 population to previously found stellar populations.
While \citet{2024A&A...682A.116N} initially associated this population with Thamnos, we find that the overlap in kinematics is small, and hence we now consider this association unlikely (\fig{fig:kinematics}). 
Among stellar populations in the halo reported to have intermediate $\xfe{Mg}$ ratio at $-1<\feh$, a stellar population dubbed Eos \citep{2022ApJ...938...21M} resembles the $\alpha$-poor 1 population most in chemical abundance and kinematics.
They identified Eos through a Gaussian mixture model (GMM) analysis of multi-dimensional chemodynamical spaces of giants with high eccentricity in APOGEE DR17 and GALAH DR3.
The existence of a similar population is confirmed by a t-SNE analysis of halo stars with high-quality spectra in APOGEE \citep{2023A&A...676A.140O}.
The average orbital energy of Eos stars is lower than Gaia-Enceladus, which again resembles the property of our $\alpha$-poor 1 population (\fig{fig:kinematics}).
We therefore consider that our $\alpha$-poor 1 population can be associated with Eos.
The sodium abundance of this population also supports its association with Eos (see \citealt{2024A&A...682A.116N}).

Nevertheless, we note that the kinematics of our $\alpha$-poor 1 population and Eos members from \citet{2022ApJ...938...21M} are slightly different (\fig{fig:kinematics}). 
The kinematic pre-selection of stars in eccentricity made by \citet{2022ApJ...938...21M} would remove any stars with high angular momenta, which explains the absence of Eos members on retrograde orbits. 
There are also Eos members at very high orbital energy, which might reflect the overlap they found between Gaia-Enceladus and Eos in the chemical space.
\citet{2022ApJ...938...21M} also include orbital energy as one of the spaces in the GMM analyses, which by design makes Eos members show a single peak in orbital energy distribution.

\citet{2022ApJ...938...21M} suggested that Eos is the population that formed in-situ after the last major merger, which would likely be the accretion of Gaia-Enceladus.
Since the sample in the present study mostly consists of main-sequence and turn-off stars, rather than APOGEE giants as in \citet{2022ApJ...938...21M}, our stars potentially provide an independent test of the scenario.
Using the ages derived by \citet{2012A&A...538A..21S}, we obtained the weighted average age of $10.4\pm0.4\,\mathrm{Gyr}$ (7 stars) for the $\alpha$-poor 1 population (Eos) and  $10.9\pm 0.8\,\mathrm{Gyr}$ (3 stars) for the $\alpha$-poor 2 population (Gaia-Enceladus).
Hence, there is no significant age difference with the current sample size and the measurement uncertainty.
If Eos formed shortly after the termination of the star formation in the progenitor of Gaia-Enceladus, this result does not contradict with the scenario presented by \citet{2022ApJ...938...21M}.

\section{Conclusion}
\label{conclusion}

We have constructed a grid of abundance corrections needed to bring 1D LTE magnesium abundances to the 3D non-LTE scale for main-sequence turn-off stars with $5000\,\mathrm{K}\lesssim \teff{} \lesssim 6500\,\mathrm{K}$, $\lgg =4.0$ and $4.5$, $-3.0 \leq \feh \leq 0.0$, and 1D LTE $\vmic{}$ of $0$, $1$, and $2\,\kms{}$.\footnote{link to CDS here}
The impact of the 3D non-LTE corrections were demonstrated on the benchmark metal-poor stars HD84937 and HD140283, and shown to significantly reduce the large ionisation imbalances found in both 1D LTE and 1D non-LTE.

The impact of the 3D non-LTE corrections were further demonstrated in the context of Galactic Archaeology, by applying them to the sample of halo stars from \citet{2010A&A...511L..10N}.
Their $\alpha$-poor population, previously interpreted as an accreted population, exhibits (at least) two distinct subpopulations in 3D non-LTE magnesium abundance. 
While the two subpopulations show different 1D LTE or 1D non-LTE magnesium abundances on average, the clear separation can be observed for the first time in 3D non-LTE thanks to the increased precision. 
Even though the sample was selected to have a narrow range in stellar parameters, the correction from 1D LTE to 3D non-LTE in $\xfe{Mg}$ varies from $-0.06\,\dex$ to $+0.08\,\dex$ with a dispersion of $0.04\,\dex$.
Therefore, the 1D LTE approximations can be the bottleneck in achieving high-precision when the spectra are of high-quality with $R>50,000$ and $S/N>100$ like those used by \citet{2010A&A...511L..10N}. 

The two $\alpha$-poor subpopulations also differ in kinematics.
While the subpopulation with the lower $\xfe{Mg}$ shows highly radial orbits, consistent with being part of Gaia-Enceladus, the other subpopulation is more tightly bound to the Milky Way's gravitational potential. 
The $\xfe{Mg}$ and kinematics as well as $\xfe{Na}$ (\citealt{2024A&A...682A.116N}) of the latter population resembles a stellar population dubbed Eos \citep{2022ApJ...938...21M}.
While the Eos was identified through a Gaussian Mixture model analysis in a chemodynamical space of a sample of stars with highly eccentric orbits, we showed that one can cleanly select Gaia-Enceladus and Eos members without kinematic preselection with precise and accurate 3D non-LTE magnesium abundance. 
Once such selections are applied to a larger sample, they will allow us to study their intrinsic distributions in, age, the present-day kinematics, and other elemental abundance ratios, with which we can reveal the formation histories of stellar populations in the Milky Way.

\begin{acknowledgements}
We thank Per J\"{o}nsson and Henrik Hartman (Malm\"{o} Universitet)
for useful discussions about the \ion{Mg}{I} oscillator strengths.
We also thank Gyuchul Myeong for sharing the lists of Eos member stars with us.
TM was supported by a Spinoza Grant from the Dutch Research Council (NWO), which was awarded to Prof. Amina Helmi, and by a Gliese Fellowship at the Zentrum f\"{u}r Astronomie, University of Heidelberg, Germany. 
AMA acknowledges support from the Swedish Research Council (VR 2020-03940).
This research was supported by computational resources provided by the
Australian Government through the National Computational Infrastructure (NCI)
under the National Computational Merit Allocation Scheme and the ANU Merit
Allocation Scheme (project y89). Some of the computations were also enabled
by resources provided by the Swedish National Infrastructure for Computing
(SNIC) at the Multidisciplinary Center for Advanced Computational Science
(UPPMAX) partially funded by the Swedish Research Council through grant
agreement no.  2018-05973.
MC acknowledges the support from the Knut and Alice Wallenberg Foundation as part of the project “Probing charge- and mass-transfer reactions on the atomic level” (2018.0028).
\end{acknowledgements}
\bibliographystyle{aa_url} 
\bibliography{bibl.bib}

\begin{appendix}
\section{Further evidence for the three distinct populations\label{appendix:A}}
This section provides a statistical confirmation of the presence of the two $\alpha$-poor subpopulations and the separation between the $\alpha$-rich and poor populations among halo stars at $-1<\feh$.
As shown in \fig{fig:MgFe1D}, the \xfe{Mg} ratio distribution among the metal-rich halo stars is best fitted with three components when Gaussian Mixture Models (GMMs) are used. 
We evaluate each GMM with a different number of components using the Bayesian information criterion (BIC), which is summarized in \tab{tab:gmmbic}.
The GMM with three components is most favoured according to the BIC. 
Notably, the BIC difference between the GMMs with two and three components is more than six, which is interpreted as a strong preference for the three-component model.
The three components correspond to the high-$\alpha$, $\alpha$-poor 1 \&2 populations.

\begin{table}
\begin{center}
\caption{BIC of GMMs \label{tab:gmmbic}}
\begin{tabular}{lr}
\hline
$N_{\rm components}$ & BIC \\\hline\hline
1 & -22.8 \\ 
2 & -24.0 \\ 
3 & -30.4 \\
4 & -12.0 \\ 
5 & -3.7  \\ \hline
\end{tabular}
\end{center}
\end{table}

\begin{figure}
\centering
\includegraphics[width=0.5\textwidth]{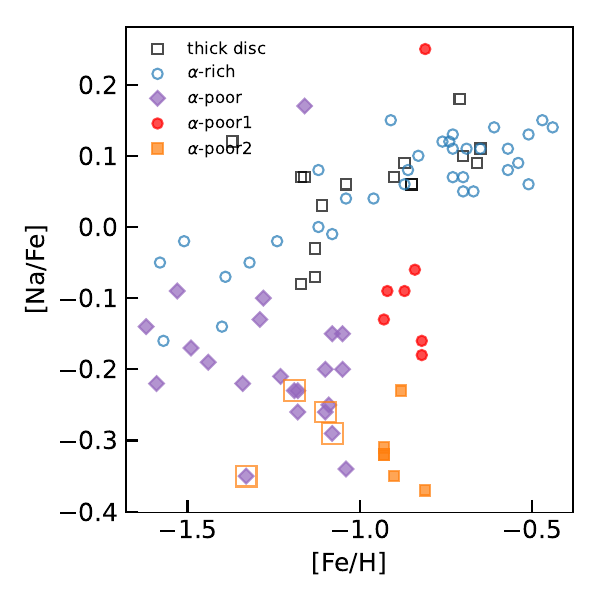}
\caption{The \xfe{Na} ratios of halo stars. The non-LTE Na abundances are taken from \citet{2024A&A...682A.116N}.\label{fig:NaFe}}
\end{figure}

We have shown in the main text that the separation between $\alpha$-poor 1\& 2 populations is also seen in kinematics. 
As indicated by \citet{2024A&A...682A.116N}, the two populations also differ in other elemental abundances, such as \xfe{Na}, which is shown in \fig{fig:NaFe}.
This reinforces our finding that $\alpha$-poor 1\& 2 populations have astrophysically different origins.

\end{appendix}

\end{document}

%% file: linelist.tex
\begin{table}
\begin{center}
\caption{Line parameters adopted in the abundance analysis of HD84937 and HD140283.}
\label{tab:linelist}
\begin{tabular}{c c c c c c c}
\hline
\hline
\noalign{\smallskip}
\multirow{1}{*}{Spec} & 
\multirow{1}{*}{$\lambda_{\text{air}} / \nm$} & 
\multirow{1}{*}{$\chi / \eV$} & 
\multirow{1}{*}{$\lggf$} & 
\multirow{1}{*}{$\log\gamma_{\text{rad}}$} & 
\multirow{1}{*}{$\sigma$} & 
\multirow{1}{*}{$\alpha$} \\ 
\noalign{\smallskip}
\hline
\hline
\noalign{\smallskip}
\ion{Mg}{I} & $    416.73$ & $    4.3458$ & $    -0.746$ & $      8.69$ & $5296$ & $0.508$ \\ 
\ion{Mg}{I} & $    457.11$ & $    0.0000$ & $    -5.732$ & $      6.00$ & $222$ & $0.253$ \\ 
\ion{Mg}{I} & $    470.30$ & $    4.3458$ & $    -0.456$ & $      8.70$ & $2827$ & $0.264$ \\ 
\ion{Mg}{I} & $    473.00$ & $    4.3458$ & $    -2.379$ & $      8.68$ & $5928$ & $0.435$ \\ 
\noalign{\smallskip}
\ion{Mg}{I} & $    516.73$ & $    2.7091$ & $    -0.854$ & $      8.02$ & $731$ & $0.240$ \\ 
\ion{Mg}{I} & $    517.27$ & $    2.7116$ & $    -0.363$ & $      8.02$ & $731$ & $0.240$ \\ 
\ion{Mg}{I} & $    518.36$ & $    2.7166$ & $    -0.168$ & $      8.02$ & $731$ & $0.240$ \\ 
\noalign{\smallskip}
\ion{Mg}{I} & $    552.84$ & $    4.3458$ & $    -0.547$ & $      8.69$ & $1461$ & $0.309$ \\ 
\ion{Mg}{I} & $    571.11$ & $    4.3458$ & $    -1.742$ & $      8.69$ & $1841$ & $0.120$ \\ 
\ion{Mg}{I} & $    880.67$ & $    4.3458$ & $    -0.144$ & $      8.69$ & $530$ & $0.277$ \\ 
\noalign{\smallskip}
\ion{Mg}{II} & $    448.11$ & $    8.8637$ & $+     0.749$ & $      8.85$ & $-$ & $-$ \\ 
\ion{Mg}{II} & $    448.12$ & $    8.8637$ & $    -0.553$ & $      8.85$ & $-$ & $-$ \\ 
\ion{Mg}{II} & $    448.13$ & $    8.8638$ & $+     0.594$ & $      8.85$ & $-$ & $-$ \\ 
\noalign{\smallskip}
\hline
\end{tabular}
\end{center}
\tablefoot{Oscillator strengths for \ion{Mg}{I} lines come from \citet{2017A&A...598A.102P}, with preference given to experimental values where available except for the $571.1\,\nm$ line (theoretical value was adopted); for the $457.1\,\nm$ line, the theoretical value from \citet{1997APS..APR.J1537J} was used instead. As discussed in the text, the radiative transfer calculations assumed $\lggf=-5.397$ for the $457.1\,\nm$ line, and $\lggf=-1.842$ for the $571.1\,\nm$. Oscillator strengths for the \ion{Mg}{II} line comes from \citet{2006ADNDT..92..607F} via NIST \citep{2020Atoms...8...56R}. ABO broadening parameters \citep[e.g.][]{2016A&ARv..24....9B} obtained via interpolating extended tables at \url{https://github.com/barklem/abo-cross} with the value for the \ion{Mg}{I} $473.0\,\nm$ line taken from the edge of the corresponding table.}
\end{table}

%% file: abundances.tex
\begin{table*}
\begin{center}
\caption{Equivalent widths $W$, reduced equivalent widths $\mathrm{REW}=\log(W/\lambda_{\mathrm{vac}})$, 3D non-LTE abundance, and abundance corrections in HD84937 and HD140283.}
\label{tab:abundances}
\begin{tabular}{c c | c c c c c | c c c c c}
\hline
\hline
\noalign{\smallskip}
\multirow{2}{*}{Spec} & 
\multirow{2}{*}{$\lambda_{\text{air}} / \nm$} & 
\multicolumn{5}{c|}{HD84937} & 
\multicolumn{5}{c}{HD140283} \\ 
& & 
\multicolumn{1}{|c}{$W/\mathrm{pm}$} & $\mathrm{REW}$ & $\lgeps{Mg}_{\text{3N}}$ & 3N-1N & 1N-1L & 
\multicolumn{1}{|c}{$W/\mathrm{pm}$} & $\mathrm{REW}$ & $\lgeps{Mg}_{\text{3N}}$ & 3N-1N & 1N-1L \\ 
\noalign{\smallskip}
\hline
\hline
\noalign{\smallskip}
\ion{Mg}{I} & $    416.73$ & 
$      3.20 \pm       0.16$ & $     -5.11$ & $      5.79 \pm       0.05$ & $+      0.08$ & $+      0.04$ & 
$      2.75 \pm       0.14$ & $     -5.18$ & $      5.45 \pm       0.04$ & $+      0.08$ & $+      0.03$ \\ 
\ion{Mg}{I} & $    457.11$ & 
$      0.35 \pm       0.04$ & $     -6.12$ & $      5.87 \pm       0.09$ & $+      0.09$ & $+      0.08$ & 
$      0.82 \pm       0.08$ & $     -5.75$ & $      5.66 \pm       0.08$ & $+      0.10$ & $+      0.08$ \\ 
\ion{Mg}{I} & $    470.30$ & 
$      4.94 \pm       0.25$ & $     -4.98$ & $      5.80 \pm       0.06$ & $+      0.11$ & $+      0.05$ & 
$      4.26 \pm       0.21$ & $     -5.04$ & $      5.44 \pm       0.04$ & $+      0.11$ & $+      0.04$ \\ 
\ion{Mg}{I} & $    473.00$ & 
$      0.14 \pm       0.01$ & $     -6.53$ & $      5.91 \pm       0.06$ & $+      0.07$ & $+      0.04$ & 
$-$ & $-$ & $-$ & $-$ & $-$ \\
\noalign{\smallskip}
\ion{Mg}{I} & $    516.73$ & 
$      10.6 \pm        0.5$ & $     -4.69$ & $      5.75 \pm       0.10$ & $+      0.17$ & $+      0.01$ & 
$      10.9 \pm        0.5$ & $     -4.68$ & $      5.35 \pm       0.10$ & $+      0.18$ & $+      0.01$ \\ 
\ion{Mg}{I} & $    517.27$ & 
$      14.8 \pm        0.7$ & $     -4.54$ & $      5.76 \pm       0.11$ & $+      0.18$ & $     -0.04$ & 
$      15.0 \pm        0.8$ & $     -4.54$ & $      5.34 \pm       0.11$ & $+      0.16$ & $-      0.02$ \\ 
\ion{Mg}{I} & $    518.36$ & 
$      16.8 \pm        0.8$ & $     -4.49$ & $      5.71 \pm       0.12$ & $+      0.16$ & $     -0.06$ & 
$      17.2 \pm        0.9$ & $     -4.48$ & $      5.32 \pm       0.10$ & $+      0.14$ & $-      0.02$ \\ 
\noalign{\smallskip}
\ion{Mg}{I} & $    552.84$ & 
$      4.69 \pm       0.23$ & $     -5.07$ & $      5.87 \pm       0.06$ & $+      0.12$ & $+      0.03$ & 
$      4.05 \pm       0.20$ & $     -5.14$ & $      5.49 \pm       0.05$ & $+      0.11$ & $+      0.02$ \\ 
\ion{Mg}{I} & $    571.11$ & 
$      0.48 \pm       0.05$ & $     -6.08$ & $      5.80 \pm       0.06$ & $+      0.08$ & $+      0.03$ & 
$      0.40 \pm       0.04$ & $     -6.15$ & $      5.47 \pm       0.05$ & $+      0.07$ & $+      0.02$ \\ 
\ion{Mg}{I} & $    880.67$ & 
$      8.05 \pm       0.40$ & $     -5.04$ & $      5.85 \pm       0.08$ & $+      0.16$ & $     -0.02$ & 
$      7.49 \pm       0.37$ & $     -5.07$ & $      5.48 \pm       0.06$ & $+      0.14$ & $-      0.00$ \\ 
\noalign{\smallskip}
\ion{Mg}{II} & $    448.12$ & 
$      3.08 \pm       0.15$ & $     -5.16$ & $      5.85 \pm       0.07$ & $     -0.01$ & $     -0.01$ & 
$      1.25 \pm       0.06$ & $     -5.55$ & $      5.55 \pm       0.05$ & $-      0.05$ & $-      0.01$ \\ 
\noalign{\smallskip}
\hline
\end{tabular}
\end{center}
\tablefoot{Three components of the \ion{Mg}{II} $448.1\,\nm$ treated together in the abundance analysis. Reported equivalent widths for this feature already exclude the contributions of the \ion{Ti}{I} blend, amounting to $0.15\,\mathrm{pm}$ and $0.12\,\mathrm{pm}$ in HD84937 and HD140283 respectively.  Uncertainties in abundances fold in uncertainties in $W$ as well as in stellar parameters.}
\end{table*}

%% file: paper_clean.bbl
\begin{thebibliography}{58}
\expandafter\ifx\csname natexlab\endcsname\relax\def\natexlab#1{#1}\fi

\bibitem[{{Abdurro'uf} {et~al.}(2022){Abdurro'uf}, {Accetta}, {Aerts}, {Silva
  Aguirre}, {Ahumada}, {Ajgaonkar}, {Filiz Ak}, {Alam}, {Allende Prieto},
  {Almeida}, {Anders}, {Anderson}, {Andrews}, {Anguiano}, {Aquino-Ort{\'\i}z},
  {Arag{\'o}n-Salamanca}, {Argudo-Fern{\'a}ndez}, {Ata}, {Aubert},
  {Avila-Reese}, {Badenes}, {Barb{\'a}}, {Barger}, {Barrera-Ballesteros},
  {Beaton}, {Beers}, {Belfiore}, {Bender}, {Bernardi}, {Bershady}, {Beutler},
  {Bidin}, {Bird}, {Bizyaev}, {Blanc}, {Blanton}, {Boardman}, {Bolton},
  {Boquien}, {Borissova}, {Bovy}, {Brandt}, {Brown}, {Brownstein}, {Brusa},
  {Buchner}, {Bundy}, {Burchett}, {Bureau}, {Burgasser}, {Cabang}, {Campbell},
  {Cappellari}, {Carlberg}, {Wanderley}, {Carrera}, {Cash}, {Chen}, {Chen},
  {Cherinka}, {Chiappini}, {Choi}, {Chojnowski}, {Chung}, {Clerc}, {Cohen},
  {Comerford}, {Comparat}, {da Costa}, {Covey}, {Crane}, {Cruz-Gonzalez},
  {Culhane}, {Cunha}, {Dai}, {Damke}, {Darling}, {Davidson}, {Davies},
  {Dawson}, {De Lee}, {Diamond-Stanic}, {Cano-D{\'\i}az}, {S{\'a}nchez},
  {Donor}, {Duckworth}, {Dwelly}, {Eisenstein}, {Elsworth}, {Emsellem},
  {Eracleous}, {Escoffier}, {Fan}, {Farr}, {Feng}, {Fern{\'a}ndez-Trincado},
  {Feuillet}, {Filipp}, {Fillingham}, {Frinchaboy}, {Fromenteau}, {Galbany},
  {Garc{\'\i}a}, {Garc{\'\i}a-Hern{\'a}ndez}, {Ge}, {Geisler}, {Gelfand},
  {G{\'e}ron}, {Gibson}, {Goddy}, {Godoy-Rivera}, {Grabowski}, {Green},
  {Greener}, {Grier}, {Griffith}, {Guo}, {Guy}, {Hadjara}, {Harding},
  {Hasselquist}, {Hayes}, {Hearty}, {Hern{\'a}ndez}, {Hill}, {Hogg},
  {Holtzman}, {Horta}, {Hsieh}, {Hsu}, {Hsu}, {Huber}, {Huertas-Company},
  {Hutchinson}, {Hwang}, {Ibarra-Medel}, {Chitham}, {Ilha}, {Imig}, {Jaekle},
  {Jayasinghe}, {Ji}, {Johnson}, {Jones}, {J{\"o}nsson}, {Katkov}, {Khalatyan},
  {Kinemuchi}, {Kisku}, {Knapen}, {Kneib}, {Kollmeier}, {Kong}, {Kounkel},
  {Kreckel}, {Krishnarao}, {Lacerna}, {Lane}, {Langgin}, {Lavender}, {Law},
  {Lazarz}, {Leung}, {Leung}, {Lewis}, {Li}, {Li}, {Lian}, {Liang}, {Lin},
  {Lin}, {Lin}, {Lintott}, {Long}, {Longa-Pe{\~n}a}, {L{\'o}pez-Cob{\'a}},
  {Lu}, {Lundgren}, {Luo}, {Mackereth}, {de la Macorra}, {Mahadevan},
  {Majewski}, {Manchado}, {Mandeville}, {Maraston}, {Margalef-Bentabol},
  {Masseron}, {Masters}, {Mathur}, {McDermid}, {Mckay}, {Merloni},
  {Merrifield}, {Meszaros}, {Miglio}, {Di Mille}, {Minniti}, {Minsley},
  {Monachesi}, {Moon}, {Mosser}, {Mulchaey}, {Muna}, {Mu{\~n}oz}, {Myers},
  {Myers}, {Nadathur}, {Nair}, {Nandra}, {Neumann}, {Newman}, {Nidever},
  {Nikakhtar}, {Nitschelm}, {O'Connell}, {Garma-Oehmichen}, {Luan Souza de
  Oliveira}, {Olney}, {Oravetz}, {Ortigoza-Urdaneta}, {Osorio}, {Otter},
  {Pace}, {Padilla}, {Pan}, {Pan}, {Parikh}, {Parker}, {Peirani}, {Pe{\~n}a
  Ram{\'\i}rez}, {Penny}, {Percival}, {Perez-Fournon}, {Pinsonneault},
  {Poidevin}, {Poovelil}, {Price-Whelan}, {B{\'a}rbara de Andrade Queiroz},
  {Raddick}, {Ray}, {Rembold}, {Riddle}, {Riffel}, {Riffel}, {Rix}, {Robin},
  {Rodr{\'\i}guez-Puebla}, {Roman-Lopes}, {Rom{\'a}n-Z{\'u}{\~n}iga}, {Rose},
  {Ross}, {Rossi}, {Rubin}, {Salvato}, {S{\'a}nchez}, {S{\'a}nchez-Gallego},
  {Sanderson}, {Santana Rojas}, {Sarceno}, {Sarmiento}, {Sayres}, {Sazonova},
  {Schaefer}, {Schiavon}, {Schlegel}, {Schneider}, {Schultheis}, {Schwope},
  {Serenelli}, {Serna}, {Shao}, {Shapiro}, {Sharma}, {Shen}, {Shetrone}, {Shu},
  {Simon}, {Skrutskie}, {Smethurst}, {Smith}, {Sobeck}, {Spoo}, {Sprague},
  {Stark}, {Stassun}, {Steinmetz}, {Stello}, {Stone-Martinez},
  {Storchi-Bergmann}, {Stringfellow}, {Stutz}, {Su}, {Taghizadeh-Popp},
  {Talbot}, {Tayar}, {Telles}, {Teske}, {Thakar}, {Theissen}, {Tkachenko},
  {Thomas}, {Tojeiro}, {Hernandez Toledo}, {Troup}, {Trump}, {Trussler},
  {Turner}, {Tuttle}, {Unda-Sanzana}, {V{\'a}zquez-Mata}, {Valentini},
  {Valenzuela}, {Vargas-Gonz{\'a}lez}, {Vargas-Maga{\~n}a}, {Alfaro},
  {Villanova}, {Vincenzo}, {Wake}, {Warfield}, {Washington}, {Weaver},
  {Weijmans}, {Weinberg}, {Weiss}, {Westfall}, {Wild}, {Wilde}, {Wilson},
  {Wilson}, {Wilson}, {Wolf}, {Wood-Vasey}, {Yan}, {Zamora}, {Zasowski},
  {Zhang}, {Zhao}, {Zheng}, {Zheng}, \& {Zhu}}]{2022ApJS..259...35A}
{Abdurro'uf}, {Accetta}, K., {Aerts}, C., {et~al.} 2022,
  \href{http://dx.doi.org/10.3847/1538-4365/ac4414}{\color{blue}\apjs},
  \href{https://ui.adsabs.harvard.edu/abs/2022ApJS..259...35A}{259, 35}

\bibitem[{{Alexeeva} {et~al.}(2018){Alexeeva}, {Ryabchikova}, {Mashonkina}, \&
  {Hu}}]{2018ApJ...866..153A}
{Alexeeva}, S., {Ryabchikova}, T., {Mashonkina}, L., \& {Hu}, S. 2018,
  \href{http://dx.doi.org/10.3847/1538-4357/aae1a8}{\color{blue}\apj},
  \href{https://ui.adsabs.harvard.edu/abs/2018ApJ...866..153A}{866, 153}

\bibitem[{{Altrock} \& {Cannon}(1975)}]{1975SoPh...42..289A}
{Altrock}, R.~C. \& {Cannon}, C.~J. 1975,
  \href{http://dx.doi.org/10.1007/BF00149912}{\color{blue}\solphys},
  \href{https://ui.adsabs.harvard.edu/abs/1975SoPh...42..289A}{42, 289}

\bibitem[{{Amarsi} {et~al.}(2022){Amarsi}, {Liljegren}, \&
  {Nissen}}]{2022A&A...668A..68A}
{Amarsi}, A.~M., {Liljegren}, S., \& {Nissen}, P.~E. 2022,
  \href{http://dx.doi.org/10.1051/0004-6361/202244542}{\color{blue}\aap},
  \href{https://ui.adsabs.harvard.edu/abs/2022A&A...668A..68A}{668, A68}

\bibitem[{{Amarsi} {et~al.}(2016){Amarsi}, {Lind}, {Asplund}, {Barklem}, \&
  {Collet}}]{2016MNRAS.463.1518A}
{Amarsi}, A.~M., {Lind}, K., {Asplund}, M., {Barklem}, P.~S., \& {Collet}, R.
  2016, \href{http://dx.doi.org/10.1093/mnras/stw2077}{\color{blue}\mnras},
  \href{http://adsabs.harvard.edu/abs/2016MNRAS.463.1518A}{463, 1518}

\bibitem[{{Amarsi} {et~al.}(2020){Amarsi}, {Lind}, {Osorio}, {Nordlander},
  {Bergemann}, {Reggiani}, {Wang}, {Buder}, {Asplund}, {Barklem}, {Wehrhahn},
  {Sk{\'u}lad{\'o}ttir}, {Kobayashi}, {Karakas}, {Gao}, {Bland-Hawthorn}, {de
  Silva}, {Kos}, {Lewis}, {Martell}, {Sharma}, {Simpson}, {Zucker},
  {{\v{C}}otar}, {Horner}, \& {Galah Collaboration}}]{2020A&A...642A..62A}
{Amarsi}, A.~M., {Lind}, K., {Osorio}, Y., {et~al.} 2020,
  \href{http://dx.doi.org/10.1051/0004-6361/202038650}{\color{blue}\aap},
  \href{https://ui.adsabs.harvard.edu/abs/2020A&A...642A..62A}{642, A62}

\bibitem[{{Amarsi} {et~al.}(2019){Amarsi}, {Nissen}, \&
  {Sk{\'u}lad{\'o}ttir}}]{2019A&A...630A.104A}
{Amarsi}, A.~M., {Nissen}, P.~E., \& {Sk{\'u}lad{\'o}ttir}, {\'A}. 2019,
  \href{http://dx.doi.org/10.1051/0004-6361/201936265}{\color{blue}\aap},
  \href{https://ui.adsabs.harvard.edu/abs/2019A&A...630A.104A}{630, A104}

\bibitem[{{Amarsi} {et~al.}(2018){Amarsi}, {Nordlander}, {Barklem}, {Asplund},
  {Collet}, \& {Lind}}]{2018A&A...615A.139A}
{Amarsi}, A.~M., {Nordlander}, T., {Barklem}, P.~S., {et~al.} 2018,
  \href{http://dx.doi.org/10.1051/0004-6361/201732546}{\color{blue}\aap},
  \href{http://adsabs.harvard.edu/abs/2018A%26A...615A.139A}{615, A139}

\bibitem[{{Asplund} {et~al.}(2021){Asplund}, {Amarsi}, \&
  {Grevesse}}]{2021A&A...653A.141A}
{Asplund}, M., {Amarsi}, A.~M., \& {Grevesse}, N. 2021,
  \href{http://dx.doi.org/10.1051/0004-6361/202140445}{\color{blue}\aap},
  \href{https://ui.adsabs.harvard.edu/abs/2021A&A...653A.141A}{653, A141}

\bibitem[{{Asplund} {et~al.}(2009){Asplund}, {Grevesse}, {Sauval}, \&
  {Scott}}]{2009ARA&A..47..481A}
{Asplund}, M., {Grevesse}, N., {Sauval}, A.~J., \& {Scott}, P. 2009,
  \href{http://dx.doi.org/10.1146/annurev.astro.46.060407.145222}{\color{blue}\araa},
  \href{http://adsabs.harvard.edu/abs/2009ARA%26A..47..481A}{47, 481}

\bibitem[{{Babusiaux} {et~al.}(2023){Babusiaux}, {Fabricius}, {Khanna},
  {Muraveva}, {Reyl{\'e}}, {Spoto}, {Vallenari}, {Luri}, {Arenou},
  {{\'A}lvarez}, {Anders}, {Antoja}, {Balbinot}, {Barache}, {Bauchet},
  {Bossini}, {Busonero}, {Cantat-Gaudin}, {Carrasco}, {Dafonte}, {Diakit{\'e}},
  {Figueras}, {Garcia-Gutierrez}, {Garofalo}, {Helmi}, {Jim{\'e}nez-Arranz},
  {Jordi}, {Kervella}, {Kostrzewa-Rutkowska}, {Leclerc}, {Licata}, {Manteiga},
  {Masip}, {Mongui{\'o}}, {Ramos}, {Robichon}, {Robin}, {Romero-G{\'o}mez},
  {S{\'a}ez}, {Santove{\~n}a}, {Spina}, {Torralba Elipe}, \&
  {Weiler}}]{2023A&A...674A..32B}
{Babusiaux}, C., {Fabricius}, C., {Khanna}, S., {et~al.} 2023,
  \href{http://dx.doi.org/10.1051/0004-6361/202243790}{\color{blue}\aap},
  \href{https://ui.adsabs.harvard.edu/abs/2023A&A...674A..32B}{674, A32}

\bibitem[{{Barklem}(2016{\natexlab{a}})}]{2016A&ARv..24....9B}
{Barklem}, P.~S. 2016{\natexlab{a}},
  \href{http://dx.doi.org/10.1007/s00159-016-0095-9}{\color{blue}\aapr},
  \href{http://adsabs.harvard.edu/abs/2016A%26ARv..24....9B}{24, 9}

\bibitem[{{Barklem}(2016{\natexlab{b}})}]{2016PhRvA..93d2705B}
{Barklem}, P.~S. 2016{\natexlab{b}},
  \href{http://dx.doi.org/10.1103/PhysRevA.93.042705}{\color{blue}\pra},
  \href{http://adsabs.harvard.edu/abs/2016PhRvA..93d2705B}{93, 042705}

\bibitem[{{Belokurov} {et~al.}(2018){Belokurov}, {Erkal}, {Evans}, {Koposov},
  \& {Deason}}]{2018MNRAS.478..611B}
{Belokurov}, V., {Erkal}, D., {Evans}, N.~W., {Koposov}, S.~E., \& {Deason},
  A.~J. 2018,
  \href{http://dx.doi.org/10.1093/mnras/sty982}{\color{blue}\mnras},
  \href{https://ui.adsabs.harvard.edu/abs/2018MNRAS.478..611B}{478, 611}

\bibitem[{{Belyaev} \& {Yakovleva}(2017)}]{2017A&A...606A.147B}
{Belyaev}, A.~K. \& {Yakovleva}, S.~A. 2017,
  \href{http://dx.doi.org/10.1051/0004-6361/201731015}{\color{blue}\aap},
  \href{http://adsabs.harvard.edu/abs/2017A%26A...606A.147B}{606, A147}

\bibitem[{{Bergemann} {et~al.}(2017){Bergemann}, {Collet}, {Amarsi}, {Kovalev},
  {Ruchti}, \& {Magic}}]{2017ApJ...847...15B}
{Bergemann}, M., {Collet}, R., {Amarsi}, A.~M., {et~al.} 2017,
  \href{http://dx.doi.org/10.3847/1538-4357/aa88cb}{\color{blue}\apj},
  \href{http://adsabs.harvard.edu/abs/2017ApJ...847...15B}{847, 15}

\bibitem[{{Buder} {et~al.}(2021){Buder}, {Sharma}, {Kos}, {Amarsi},
  {Nordlander}, {Lind}, {Martell}, {Asplund}, {Bland-Hawthorn}, {Casey}, {de
  Silva}, {D'Orazi}, {Freeman}, {Hayden}, {Lewis}, {Lin}, {Schlesinger},
  {Simpson}, {Stello}, {Zucker}, {Zwitter}, {Beeson}, {Buck}, {Casagrande},
  {Clark}, {{\v{C}}otar}, {da Costa}, {de Grijs}, {Feuillet}, {Horner},
  {Kafle}, {Khanna}, {Kobayashi}, {Liu}, {Montet}, {Nandakumar}, {Nataf},
  {Ness}, {Spina}, {Tepper-Garc{\'\i}a}, {Ting}, {Traven},
  {Vogrin{\v{c}}i{\v{c}}}, {Wittenmyer}, {Wyse}, {{\v{Z}}erjal}, \& {GALAH
  Collaboration}}]{2021MNRAS.506..150B}
{Buder}, S., {Sharma}, S., {Kos}, J., {et~al.} 2021,
  \href{http://dx.doi.org/10.1093/mnras/stab1242}{\color{blue}\mnras},
  \href{https://ui.adsabs.harvard.edu/abs/2021MNRAS.506..150B}{506, 150}

\bibitem[{{Casagrande} {et~al.}(2010){Casagrande}, {Ram{\'{\i}}rez},
  {Mel{\'e}ndez}, {Bessell}, \& {Asplund}}]{2010A&A...512A..54C}
{Casagrande}, L., {Ram{\'{\i}}rez}, I., {Mel{\'e}ndez}, J., {Bessell}, M., \&
  {Asplund}, M. 2010,
  \href{http://dx.doi.org/10.1051/0004-6361/200913204}{\color{blue}\aap},
  \href{http://adsabs.harvard.edu/abs/2010A%26A...512A..54C}{512, A54}

\bibitem[{{Chang} \& {Tang}(1990)}]{1990JQSRT..43..207C}
{Chang}, T.~N. \& {Tang}, X. 1990,
  \href{http://dx.doi.org/10.1016/0022-4073(90)90053-9}{\color{blue}\jqsrt},
  \href{https://ui.adsabs.harvard.edu/abs/1990JQSRT..43..207C}{43, 207}

\bibitem[{{Dodd} {et~al.}(2023){Dodd}, {Callingham}, {Helmi}, {Matsuno},
  {Ruiz-Lara}, {Balbinot}, \& {L{\"o}vdal}}]{2023A&A...670L...2D}
{Dodd}, E., {Callingham}, T.~M., {Helmi}, A., {et~al.} 2023,
  \href{http://dx.doi.org/10.1051/0004-6361/202244546}{\color{blue}\aap},
  \href{https://ui.adsabs.harvard.edu/abs/2023A&A...670L...2D}{670, L2}

\bibitem[{{Froese Fischer} {et~al.}(2006){Froese Fischer}, {Tachiev}, \&
  {Irimia}}]{2006ADNDT..92..607F}
{Froese Fischer}, C., {Tachiev}, G., \& {Irimia}, A. 2006,
  \href{http://dx.doi.org/10.1016/j.adt.2006.03.001}{\color{blue}\adndt},
  \href{https://ui.adsabs.harvard.edu/abs/2006ADNDT..92..607F}{92, 607}

\bibitem[{{Fuhrmann}(1998)}]{1998A&A...338..161F}
{Fuhrmann}, K. 1998, \aap,
  \href{https://ui.adsabs.harvard.edu/abs/1998A&A...338..161F}{338, 161}

\bibitem[{{Gaia Collaboration} {et~al.}(2023){Gaia Collaboration}, {Vallenari},
  {Brown}, {Prusti}, {de Bruijne}, {Arenou}, {Babusiaux}, {Biermann},
  {Creevey}, {Ducourant}, {Evans}, {Eyer}, {Guerra}, {Hutton}, {Jordi},
  {Klioner}, {Lammers}, {Lindegren}, {Luri}, {Mignard}, {Panem}, {Pourbaix},
  {Randich}, {Sartoretti}, {Soubiran}, {Tanga}, {Walton}, {Bailer-Jones},
  {Bastian}, {Drimmel}, {Jansen}, {Katz}, {Lattanzi}, {van Leeuwen}, {Bakker},
  {Cacciari}, {Casta{\~n}eda}, {De Angeli}, {Fabricius}, {Fouesneau},
  {Fr{\'e}mat}, {Galluccio}, {Guerrier}, {Heiter}, {Masana}, {Messineo},
  {Mowlavi}, {Nicolas}, {Nienartowicz}, {Pailler}, {Panuzzo}, {Riclet}, {Roux},
  {Seabroke}, {Sordo}, {Th{\'e}venin}, {Gracia-Abril}, {Portell}, {Teyssier},
  {Altmann}, {Andrae}, {Audard}, {Bellas-Velidis}, {Benson}, {Berthier},
  {Blomme}, {Burgess}, {Busonero}, {Busso}, {C{\'a}novas}, {Carry}, {Cellino},
  {Cheek}, {Clementini}, {Damerdji}, {Davidson}, {de Teodoro}, {Nu{\~n}ez
  Campos}, {Delchambre}, {Dell'Oro}, {Esquej}, {Fern{\'a}ndez-Hern{\'a}ndez},
  {Fraile}, {Garabato}, {Garc{\'\i}a-Lario}, {Gosset}, {Haigron}, {Halbwachs},
  {Hambly}, {Harrison}, {Hern{\'a}ndez}, {Hestroffer}, {Hodgkin}, {Holl},
  {Jan{\ss}en}, {Jevardat de Fombelle}, {Jordan}, {Krone-Martins}, {Lanzafame},
  {L{\"o}ffler}, {Marchal}, {Marrese}, {Moitinho}, {Muinonen}, {Osborne},
  {Pancino}, {Pauwels}, {Recio-Blanco}, {Reyl{\'e}}, {Riello}, {Rimoldini},
  {Roegiers}, {Rybizki}, {Sarro}, {Siopis}, {Smith}, {Sozzetti}, {Utrilla},
  {van Leeuwen}, {Abbas}, {{\'A}brah{\'a}m}, {Abreu Aramburu}, {Aerts},
  {Aguado}, {Ajaj}, {Aldea-Montero}, {Altavilla}, {{\'A}lvarez}, {Alves},
  {Anders}, {Anderson}, {Anglada Varela}, {Antoja}, {Baines}, {Baker},
  {Balaguer-N{\'u}{\~n}ez}, {Balbinot}, {Balog}, {Barache}, {Barbato},
  {Barros}, {Barstow}, {Bartolom{\'e}}, {Bassilana}, {Bauchet}, {Becciani},
  {Bellazzini}, {Berihuete}, {Bernet}, {Bertone}, {Bianchi}, {Binnenfeld},
  {Blanco-Cuaresma}, {Blazere}, {Boch}, {Bombrun}, {Bossini}, {Bouquillon},
  {Bragaglia}, {Bramante}, {Breedt}, {Bressan}, {Brouillet}, {Brugaletta},
  {Bucciarelli}, {Burlacu}, {Butkevich}, {Buzzi}, {Caffau}, {Cancelliere},
  {Cantat-Gaudin}, {Carballo}, {Carlucci}, {Carnerero}, {Carrasco},
  {Casamiquela}, {Castellani}, {Castro-Ginard}, {Chaoul}, {Charlot}, {Chemin},
  {Chiaramida}, {Chiavassa}, {Chornay}, {Comoretto}, {Contursi}, {Cooper},
  {Cornez}, {Cowell}, {Crifo}, {Cropper}, {Crosta}, {Crowley}, {Dafonte},
  {Dapergolas}, {David}, {David}, {de Laverny}, {De Luise}, {De March}, {De
  Ridder}, {de Souza}, {de Torres}, {del Peloso}, {del Pozo}, {Delbo},
  {Delgado}, {Delisle}, {Demouchy}, {Dharmawardena}, {Di Matteo}, {Diakite},
  {Diener}, {Distefano}, {Dolding}, {Edvardsson}, {Enke}, {Fabre}, {Fabrizio},
  {Faigler}, {Fedorets}, {Fernique}, {Fienga}, {Figueras}, {Fournier},
  {Fouron}, {Fragkoudi}, {Gai}, {Garcia-Gutierrez}, {Garcia-Reinaldos},
  {Garc{\'\i}a-Torres}, {Garofalo}, {Gavel}, {Gavras}, {Gerlach}, {Geyer},
  {Giacobbe}, {Gilmore}, {Girona}, {Giuffrida}, {Gomel}, {Gomez},
  {Gonz{\'a}lez-N{\'u}{\~n}ez}, {Gonz{\'a}lez-Santamar{\'\i}a},
  {Gonz{\'a}lez-Vidal}, {Granvik}, {Guillout}, {Guiraud},
  {Guti{\'e}rrez-S{\'a}nchez}, {Guy}, {Hatzidimitriou}, {Hauser}, {Haywood},
  {Helmer}, {Helmi}, {Sarmiento}, {Hidalgo}, {Hilger}, {H{\l}adczuk}, {Hobbs},
  {Holland}, {Huckle}, {Jardine}, {Jasniewicz}, {Jean-Antoine Piccolo},
  {Jim{\'e}nez-Arranz}, {Jorissen}, {Juaristi Campillo}, {Julbe}, {Karbevska},
  {Kervella}, {Khanna}, {Kontizas}, {Kordopatis}, {Korn}, {K{\'o}sp{\'a}l},
  {Kostrzewa-Rutkowska}, {Kruszy{\'n}ska}, {Kun}, {Laizeau}, {Lambert},
  {Lanza}, {Lasne}, {Le Campion}, {Lebreton}, {Lebzelter}, {Leccia}, {Leclerc},
  {Lecoeur-Taibi}, {Liao}, {Licata}, {Lindstr{\o}m}, {Lister}, {Livanou},
  {Lobel}, {Lorca}, {Loup}, {Madrero Pardo}, {Magdaleno Romeo}, {Managau},
  {Mann}, {Manteiga}, {Marchant}, {Marconi}, {Marcos}, {Marcos Santos},
  {Mar{\'\i}n Pina}, {Marinoni}, {Marocco}, {Marshall}, {Martin Polo},
  {Mart{\'\i}n-Fleitas}, {Marton}, {Mary}, {Masip}, {Massari},
  {Mastrobuono-Battisti}, {Mazeh}, {McMillan}, {Messina}, {Michalik}, {Millar},
  {Mints}, {Molina}, {Molinaro}, {Moln{\'a}r}, {Monari}, {Mongui{\'o}},
  {Montegriffo}, {Montero}, {Mor}, {Mora}, {Morbidelli}, {Morel}, {Morris},
  {Muraveva}, {Murphy}, {Musella}, {Nagy}, {Noval}, {Oca{\~n}a}, {Ogden},
  {Ordenovic}, {Osinde}, {Pagani}, {Pagano}, {Palaversa}, {Palicio},
  {Pallas-Quintela}, {Panahi}, {Payne-Wardenaar}, {Pe{\~n}alosa Esteller},
  {Penttil{\"a}}, {Pichon}, {Piersimoni}, {Pineau}, {Plachy}, {Plum}, {Poggio},
  {Pr{\v{s}}a}, {Pulone}, {Racero}, {Ragaini}, {Rainer}, {Raiteri}, {Rambaux},
  {Ramos}, {Ramos-Lerate}, {Re Fiorentin}, {Regibo}, {Richards}, {Rios Diaz},
  {Ripepi}, {Riva}, {Rix}, {Rixon}, {Robichon}, {Robin}, {Robin}, {Roelens},
  {Rogues}, {Rohrbasser}, {Romero-G{\'o}mez}, {Rowell}, {Royer}, {Ruz Mieres},
  {Rybicki}, {Sadowski}, {S{\'a}ez N{\'u}{\~n}ez}, {Sagrist{\`a} Sell{\'e}s},
  {Sahlmann}, {Salguero}, {Samaras}, {Sanchez Gimenez}, {Sanna},
  {Santove{\~n}a}, {Sarasso}, {Schultheis}, {Sciacca}, {Segol}, {Segovia},
  {S{\'e}gransan}, {Semeux}, {Shahaf}, {Siddiqui}, {Siebert}, {Siltala},
  {Silvelo}, {Slezak}, {Slezak}, {Smart}, {Snaith}, {Solano}, {Solitro},
  {Souami}, {Souchay}, {Spagna}, {Spina}, {Spoto}, {Steele},
  {Steidelm{\"u}ller}, {Stephenson}, {S{\"u}veges}, {Surdej}, {Szabados},
  {Szegedi-Elek}, {Taris}, {Taylor}, {Teixeira}, {Tolomei}, {Tonello}, {Torra},
  {Torra}, {Torralba Elipe}, {Trabucchi}, {Tsounis}, {Turon}, {Ulla}, {Unger},
  {Vaillant}, {van Dillen}, {van Reeven}, {Vanel}, {Vecchiato}, {Viala},
  {Vicente}, {Voutsinas}, {Weiler}, {Wevers}, {Wyrzykowski}, {Yoldas}, {Yvard},
  {Zhao}, {Zorec}, {Zucker}, \& {Zwitter}}]{2023A&A...674A...1G}
{Gaia Collaboration}, {Vallenari}, A., {Brown}, A.~G.~A., {et~al.} 2023,
  \href{http://dx.doi.org/10.1051/0004-6361/202243940}{\color{blue}\aap},
  \href{https://ui.adsabs.harvard.edu/abs/2023A&A...674A...1G}{674, A1}

\bibitem[{{Giribaldi} {et~al.}(2021){Giribaldi}, {da Silva}, {Smiljanic}, \&
  {Cornejo Espinoza}}]{2021A&A...650A.194G}
{Giribaldi}, R.~E., {da Silva}, A.~R., {Smiljanic}, R., \& {Cornejo Espinoza},
  D. 2021,
  \href{http://dx.doi.org/10.1051/0004-6361/202140751}{\color{blue}\aap},
  \href{https://ui.adsabs.harvard.edu/abs/2021A&A...650A.194G}{650, A194}

\bibitem[{{Godone} \& {Novero}(1992)}]{1992PhRvA..45.1717G}
{Godone}, A. \& {Novero}, C. 1992,
  \href{http://dx.doi.org/10.1103/PhysRevA.45.1717}{\color{blue}\pra},
  \href{https://ui.adsabs.harvard.edu/abs/1992PhRvA..45.1717G}{45, 1717}

\bibitem[{{Grevesse} {et~al.}(2007){Grevesse}, {Asplund}, \&
  {Sauval}}]{2007coma.book..105G}
{Grevesse}, N., {Asplund}, M., \& {Sauval}, A.~J. 2007, {The Solar Chemical
  Composition}, ed. R.~{von Steiger}, G.~{Gloeckler}, \& G.~M. {Mason}
  (Springer Science+Business Media), 105

\bibitem[{{Gustafsson} {et~al.}(2008){Gustafsson}, {Edvardsson}, {Eriksson},
  {J{\o}rgensen}, {Nordlund}, \& {Plez}}]{2008A&A...486..951G}
{Gustafsson}, B., {Edvardsson}, B., {Eriksson}, K., {et~al.} 2008,
  \href{http://dx.doi.org/10.1051/0004-6361:200809724}{\color{blue}\aap},
  \href{http://adsabs.harvard.edu/abs/2008A%26A...486..951G}{486, 951}

\bibitem[{{Heiter} {et~al.}(2015){Heiter}, {Jofr{\'e}}, {Gustafsson}, {Korn},
  {Soubiran}, \& {Th{\'e}venin}}]{2015A&A...582A..49H}
{Heiter}, U., {Jofr{\'e}}, P., {Gustafsson}, B., {et~al.} 2015,
  \href{http://dx.doi.org/10.1051/0004-6361/201526319}{\color{blue}\aap},
  \href{http://adsabs.harvard.edu/abs/2015A%26A...582A..49H}{582, A49}

\bibitem[{{Helmi} {et~al.}(2018){Helmi}, {Babusiaux}, {Koppelman}, {Massari},
  {Veljanoski}, \& {Brown}}]{2018Natur.563...85H}
{Helmi}, A., {Babusiaux}, C., {Koppelman}, H.~H., {et~al.} 2018,
  \href{http://dx.doi.org/10.1038/s41586-018-0625-x}{\color{blue}\nat},
  \href{http://adsabs.harvard.edu/abs/2018Natur.563...85H}{563, 85}

\bibitem[{{Horta} {et~al.}(2023){Horta}, {Schiavon}, {Mackereth}, {Weinberg},
  {Hasselquist}, {Feuillet}, {O'Connell}, {Anguiano}, {Allende-Prieto},
  {Beaton}, {Bizyaev}, {Cunha}, {Geisler}, {Garc{\'\i}a-Hern{\'a}ndez},
  {Holtzman}, {J{\"o}nsson}, {Lane}, {Majewski}, {M{\'e}sz{\'a}ros}, {Minniti},
  {Nitschelm}, {Shetrone}, {Smith}, \& {Zasowski}}]{2023MNRAS.520.5671H}
{Horta}, D., {Schiavon}, R.~P., {Mackereth}, J.~T., {et~al.} 2023,
  \href{http://dx.doi.org/10.1093/mnras/stac3179}{\color{blue}\mnras},
  \href{https://ui.adsabs.harvard.edu/abs/2023MNRAS.520.5671H}{520, 5671}

\bibitem[{{Ji} {et~al.}(2020){Ji}, {Li}, {Hansen}, {Casey}, {Koposov}, {Pace},
  {Mackey}, {Lewis}, {Simpson}, {Bland-Hawthorn}, {Cullinane}, {Da Costa},
  {Hattori}, {Martell}, {Kuehn}, {Erkal}, {Shipp}, {Wan}, \&
  {Zucker}}]{2020AJ....160..181J}
{Ji}, A.~P., {Li}, T.~S., {Hansen}, T.~T., {et~al.} 2020,
  \href{http://dx.doi.org/10.3847/1538-3881/abacb6}{\color{blue}\aj},
  \href{https://ui.adsabs.harvard.edu/abs/2020AJ....160..181J}{160, 181}

\bibitem[{{Jofr{\'e}} {et~al.}(2015){Jofr{\'e}}, {Heiter}, {Soubiran},
  {Blanco-Cuaresma}, {Masseron}, {Nordlander}, {Chemin}, {Worley}, {Van Eck},
  {Hourihane}, {Gilmore}, {Adibekyan}, {Bergemann}, {Cantat-Gaudin},
  {Delgado-Mena}, {Gonz{\'a}lez Hern{\'a}ndez}, {Guiglion}, {Lardo}, {de
  Laverny}, {Lind}, {Magrini}, {Mikolaitis}, {Montes}, {Pancino},
  {Recio-Blanco}, {Sordo}, {Sousa}, {Tabernero}, \&
  {Vallenari}}]{2015A&A...582A..81J}
{Jofr{\'e}}, P., {Heiter}, U., {Soubiran}, C., {et~al.} 2015,
  \href{http://dx.doi.org/10.1051/0004-6361/201526604}{\color{blue}\aap},
  \href{http://adsabs.harvard.edu/abs/2015A%26A...582A..81J}{582, A81}

\bibitem[{{J{\"o}nsson} \& {Froese Fischer}(1997)}]{1997APS..APR.J1537J}
{J{\"o}nsson}, P. \& {Froese Fischer}, C. 1997, in APS April Meeting Abstracts,
  APS Meeting Abstracts,
  \href{https://ui.adsabs.harvard.edu/abs/1997APS..APR.J1537J}{J15.37}

\bibitem[{{Karovicova} {et~al.}(2020){Karovicova}, {White}, {Nordlander},
  {Casagrand e}, {Ireland}, {Huber}, \& {Jofr{\'e}}}]{2020A&A...640A..25K}
{Karovicova}, I., {White}, T.~R., {Nordlander}, T., {et~al.} 2020,
  \href{http://dx.doi.org/10.1051/0004-6361/202037590}{\color{blue}\aap},
  \href{https://ui.adsabs.harvard.edu/abs/2020A&A...640A..25K}{640, A25}

\bibitem[{{Karovicova} {et~al.}(2018){Karovicova}, {White}, {Nordlander},
  {Lind}, {Casagrande}, {Ireland}, {Huber}, {Creevey}, {Mourard}, {Schaefer},
  {Gilmore}, {Chiavassa}, {Wittkowski}, {Jofr{\'e}}, {Heiter}, {Th{\'e}venin},
  \& {Asplund}}]{2018MNRAS.475L..81K}
{Karovicova}, I., {White}, T.~R., {Nordlander}, T., {et~al.} 2018,
  \href{http://dx.doi.org/10.1093/mnrasl/sly010}{\color{blue}\mnras},
  \href{https://ui.adsabs.harvard.edu/abs/2018MNRAS.475L..81K}{475, L81}

\bibitem[{{Lagae} {et~al.}(2023){Lagae}, {Amarsi}, {Rodr{\'\i}guez D{\'\i}az},
  {Lind}, {Nordlander}, {Hansen}, \& {Heger}}]{2023A&A...672A..90L}
{Lagae}, C., {Amarsi}, A.~M., {Rodr{\'\i}guez D{\'\i}az}, L.~F., {et~al.} 2023,
  \href{http://dx.doi.org/10.1051/0004-6361/202245786}{\color{blue}\aap},
  \href{https://ui.adsabs.harvard.edu/abs/2023A&A...672A..90L}{672, A90}

\bibitem[{{Lambert}(1993)}]{1993PhST...47..186L}
{Lambert}, D.~L. 1993,
  \href{http://dx.doi.org/10.1088/0031-8949/1993/T47/030}{\color{blue}Physica
  Scripta Volume T},
  \href{http://adsabs.harvard.edu/abs/1993PhST...47..186L}{47, 186}

\bibitem[{{Langangen} \& {Carlsson}(2009)}]{2009ApJ...696.1892L}
{Langangen}, {\O}. \& {Carlsson}, M. 2009,
  \href{http://dx.doi.org/10.1088/0004-637X/696/2/1892}{\color{blue}\apj},
  \href{https://ui.adsabs.harvard.edu/abs/2009ApJ...696.1892L}{696, 1892}

\bibitem[{{Lawler} {et~al.}(2013){Lawler}, {Guzman}, {Wood}, {Sneden}, \&
  {Cowan}}]{2013ApJS..205...11L}
{Lawler}, J.~E., {Guzman}, A., {Wood}, M.~P., {Sneden}, C., \& {Cowan}, J.~J.
  2013, \href{http://dx.doi.org/10.1088/0067-0049/205/2/11}{\color{blue}\apjs},
  \href{https://ui.adsabs.harvard.edu/abs/2013ApJS..205...11L}{205, 11}

\bibitem[{{Leenaarts} \& {Carlsson}(2009)}]{2009ASPC..415...87L_short}
{Leenaarts}, J. \& {Carlsson}, M. 2009, in ASP, Vol. 415, The Second Hinode
  Science Meeting, ed. B.~{Lites}, M.~{Cheung}, T.~{Magara}, J.~{Mariska}, \&
  K.~{Reeves}, \href{http://adsabs.harvard.edu/abs/2009ASPC..415...87L}{87}

\bibitem[{{Lind} {et~al.}(2022){Lind}, {Nordlander}, {Wehrhahn}, {Montelius},
  {Osorio}, {Barklem}, {Af{\c{s}}ar}, {Sneden}, \&
  {Kobayashi}}]{2022A&A...665A..33L}
{Lind}, K., {Nordlander}, T., {Wehrhahn}, A., {et~al.} 2022,
  \href{http://dx.doi.org/10.1051/0004-6361/202142195}{\color{blue}\aap},
  \href{https://ui.adsabs.harvard.edu/abs/2022A&A...665A..33L}{665, A33}

\bibitem[{{Magic} {et~al.}(2013){Magic}, {Collet}, {Asplund}, {Trampedach},
  {Hayek}, {Chiavassa}, {Stein}, \& {Nordlund}}]{2013A&A...557A..26M}
{Magic}, Z., {Collet}, R., {Asplund}, M., {et~al.} 2013,
  \href{http://dx.doi.org/10.1051/0004-6361/201321274}{\color{blue}\aap},
  \href{http://adsabs.harvard.edu/abs/2013A%26A...557A..26M}{557, A26}

\bibitem[{{Mallinson} {et~al.}(2022){Mallinson}, {Lind}, {Amarsi}, {Barklem},
  {Grumer}, {Belyaev}, \& {Youakim}}]{2022A&A...668A.103M}
{Mallinson}, J.~W.~E., {Lind}, K., {Amarsi}, A.~M., {et~al.} 2022,
  \href{http://dx.doi.org/10.1051/0004-6361/202244788}{\color{blue}\aap},
  \href{https://ui.adsabs.harvard.edu/abs/2022A&A...668A.103M}{668, A103}

\bibitem[{{Matsuno} {et~al.}(2022{\natexlab{a}}){Matsuno}, {Dodd}, {Koppelman},
  {Helmi}, {Ishigaki}, {Aoki}, {Zhao}, {Yuan}, \&
  {Hattori}}]{2022A&A...665A..46M}
{Matsuno}, T., {Dodd}, E., {Koppelman}, H.~H., {et~al.} 2022{\natexlab{a}},
  \href{http://dx.doi.org/10.1051/0004-6361/202243609}{\color{blue}\aap},
  \href{https://ui.adsabs.harvard.edu/abs/2022A&A...665A..46M}{665, A46}

\bibitem[{{Matsuno} {et~al.}(2022{\natexlab{b}}){Matsuno}, {Koppelman},
  {Helmi}, {Aoki}, {Ishigaki}, {Suda}, {Yuan}, \&
  {Hattori}}]{2022A&A...661A.103M}
{Matsuno}, T., {Koppelman}, H.~H., {Helmi}, A., {et~al.} 2022{\natexlab{b}},
  \href{http://dx.doi.org/10.1051/0004-6361/202142752}{\color{blue}\aap},
  \href{https://ui.adsabs.harvard.edu/abs/2022A&A...661A.103M}{661, A103}

\bibitem[{{Myeong} {et~al.}(2022){Myeong}, {Belokurov}, {Aguado}, {Evans},
  {Caldwell}, \& {Bradley}}]{2022ApJ...938...21M}
{Myeong}, G.~C., {Belokurov}, V., {Aguado}, D.~S., {et~al.} 2022,
  \href{http://dx.doi.org/10.3847/1538-4357/ac8d68}{\color{blue}\apj},
  \href{https://ui.adsabs.harvard.edu/abs/2022ApJ...938...21M}{938, 21}

\bibitem[{{Nissen} {et~al.}(2024){Nissen}, {Amarsi}, {Sk{\'u}lad{\'o}ttir}, \&
  {Schuster}}]{2024A&A...682A.116N}
{Nissen}, P.~E., {Amarsi}, A.~M., {Sk{\'u}lad{\'o}ttir}, {\'A}., \& {Schuster},
  W.~J. 2024,
  \href{http://dx.doi.org/10.1051/0004-6361/202348392}{\color{blue}\aap},
  \href{https://ui.adsabs.harvard.edu/abs/2024A&A...682A.116N}{682, A116}

\bibitem[{{Nissen} {et~al.}(2014){Nissen}, {Chen}, {Carigi}, {Schuster}, \&
  {Zhao}}]{2014A&A...568A..25N}
{Nissen}, P.~E., {Chen}, Y.~Q., {Carigi}, L., {Schuster}, W.~J., \& {Zhao}, G.
  2014, \href{http://dx.doi.org/10.1051/0004-6361/201424184}{\color{blue}\aap},
  \href{http://adsabs.harvard.edu/abs/2014A%26A...568A..25N}{568, A25}

\bibitem[{{Nissen} \& {Schuster}(2010)}]{2010A&A...511L..10N}
{Nissen}, P.~E. \& {Schuster}, W.~J. 2010,
  \href{http://dx.doi.org/10.1051/0004-6361/200913877}{\color{blue}\aap},
  \href{http://adsabs.harvard.edu/abs/2010A%26A...511L..10N}{511, L10}

\bibitem[{{Nissen} \& {Schuster}(2011)}]{2011A&A...530A..15N}
{Nissen}, P.~E. \& {Schuster}, W.~J. 2011,
  \href{http://dx.doi.org/10.1051/0004-6361/201116619}{\color{blue}\aap},
  \href{http://adsabs.harvard.edu/abs/2011A%26A...530A..15N}{530, A15}

\bibitem[{{Ortigoza-Urdaneta} {et~al.}(2023){Ortigoza-Urdaneta}, {Vieira},
  {Fern{\'a}ndez-Trincado}, {Queiroz}, {Barbuy}, {Beers}, {Chiappini},
  {Anders}, {Minniti}, \& {Tang}}]{2023A&A...676A.140O}
{Ortigoza-Urdaneta}, M., {Vieira}, K., {Fern{\'a}ndez-Trincado}, J.~G.,
  {et~al.} 2023,
  \href{http://dx.doi.org/10.1051/0004-6361/202346325}{\color{blue}\aap},
  \href{https://ui.adsabs.harvard.edu/abs/2023A&A...676A.140O}{676, A140}

\bibitem[{{Osorio} {et~al.}(2015){Osorio}, {Barklem}, {Lind}, {Belyaev},
  {Spielfiedel}, {Guitou}, \& {Feautrier}}]{2015A&A...579A..53O}
{Osorio}, Y., {Barklem}, P.~S., {Lind}, K., {et~al.} 2015,
  \href{http://dx.doi.org/10.1051/0004-6361/201525846}{\color{blue}\aap},
  \href{http://adsabs.harvard.edu/abs/2015A%26A...579A..53O}{579, A53}

\bibitem[{{Pehlivan Rhodin} {et~al.}(2017){Pehlivan Rhodin}, {Hartman},
  {Nilsson}, \& {J{\"o}nsson}}]{2017A&A...598A.102P}
{Pehlivan Rhodin}, A., {Hartman}, H., {Nilsson}, H., \& {J{\"o}nsson}, P. 2017,
  \href{http://dx.doi.org/10.1051/0004-6361/201629849}{\color{blue}\aap},
  \href{http://adsabs.harvard.edu/abs/2017A%26A...598A.102P}{598, A102}

\bibitem[{{Ralchenko} \& {Kramida}(2020)}]{2020Atoms...8...56R}
{Ralchenko}, Y. \& {Kramida}, A. 2020,
  \href{http://dx.doi.org/10.3390/atoms8030056}{\color{blue}Atoms},
  \href{https://ui.adsabs.harvard.edu/abs/2020Atoms...8...56R}{8, 56}

\bibitem[{{Ruiz-Lara} {et~al.}(2022){Ruiz-Lara}, {Matsuno}, {L{\"o}vdal},
  {Helmi}, {Dodd}, \& {Koppelman}}]{2022A&A...665A..58R}
{Ruiz-Lara}, T., {Matsuno}, T., {L{\"o}vdal}, S.~S., {et~al.} 2022,
  \href{http://dx.doi.org/10.1051/0004-6361/202243061}{\color{blue}\aap},
  \href{https://ui.adsabs.harvard.edu/abs/2022A&A...665A..58R}{665, A58}

\bibitem[{{Schuster} {et~al.}(2012){Schuster}, {Moreno}, {Nissen}, \&
  {Pichardo}}]{2012A&A...538A..21S}
{Schuster}, W.~J., {Moreno}, E., {Nissen}, P.~E., \& {Pichardo}, B. 2012,
  \href{http://dx.doi.org/10.1051/0004-6361/201118035}{\color{blue}\aap},
  \href{https://ui.adsabs.harvard.edu/abs/2012A%26A...538A..21S}{538, A21}

\bibitem[{{Weinberg} {et~al.}(2019){Weinberg}, {Holtzman}, {Hasselquist},
  {Bird}, {Johnson}, {Shetrone}, {Sobeck}, {Allende Prieto}, {Bizyaev},
  {Carrera}, {Cohen}, {Cunha}, {Ebelke}, {Fernandez-Trincado},
  {Garc{\'\i}a-Hern{\'a}ndez}, {Hayes}, {J{\"o}nsson}, {Lane}, {Majewski},
  {Malanushenko}, {M{\'e}sz{\'a}ros}, {Nidever}, {Nitschelm}, {Pan}, {Rix},
  {Rybizki}, {Schiavon}, {Schneider}, {Wilson}, \&
  {Zamora}}]{2019ApJ...874..102W}
{Weinberg}, D.~H., {Holtzman}, J.~A., {Hasselquist}, S., {et~al.} 2019,
  \href{http://dx.doi.org/10.3847/1538-4357/ab07c7}{\color{blue}\apj},
  \href{https://ui.adsabs.harvard.edu/abs/2019ApJ...874..102W}{874, 102}

\bibitem[{{Zhou} {et~al.}(2023){Zhou}, {Amarsi}, {Aguirre B{\o}rsen-Koch},
  {Karlsmose}, {Collet}, \& {Nordlander}}]{2023A&A...677A..98Z}
{Zhou}, Y., {Amarsi}, A.~M., {Aguirre B{\o}rsen-Koch}, V., {et~al.} 2023,
  \href{http://dx.doi.org/10.1051/0004-6361/202346398}{\color{blue}\aap},
  \href{https://ui.adsabs.harvard.edu/abs/2023A&A...677A..98Z}{677, A98}

\end{thebibliography}
